\documentclass[journal]{IEEEtran}

\usepackage{standalone,placeins}
\usepackage{mathptmx,mathtools,mathrsfs}
\usepackage{amsmath,amssymb,amsfonts,amsthm}
\usepackage{graphicx,caption,subcaption}
\usepackage{textcomp,multirow,multicol}
\usepackage{hyperref}
\usepackage{booktabs}
\usepackage{tikz}
\usepackage{pgfplots}
\usepackage{xcolor}
\usepackage{colortbl}
\usepackage{algpseudocode}
\usepackage{algorithm}
\usepackage{cite}

\usepgfplotslibrary{groupplots}
\pgfplotsset{compat=newest}

\pgfplotsset{every axis/.append style={
		label style={font=\small},
		tick label style={font=\small}  
}}

\usetikzlibrary{decorations.markings}
\usetikzlibrary{datavisualization.formats.functions,arrows.meta}

\definecolor{NL1}{rgb}{1.00, 0.00, 0.00}
\definecolor{NL5}{rgb}{0.75, 0.00, 0.20}
\definecolor{RRMP}{rgb}{1.00, 0.50, 0.00}
\definecolor{RBMP}{rgb}{1.00, 0.80, 0.00}
\definecolor{SRMP}{rgb}{0.40, 0.69, 0.20}
\definecolor{SBMP}{rgb}{0.13, 0.55, 0.13}
\definecolor{TRMP}{rgb}{0.00, 0.28, 0.67}
\definecolor{TBMP}{rgb}{0.29, 0.59, 0.82}

\newcommand{\gl}{g_{\rm{MMPS}}}
\newcommand{\Rl}{\mathcal{R}_{\rm{MMPS}}}
\newcommand{\gq}{g_{\rm{ELPS}}}
\newcommand{\Rq}{\mathcal{R}_{\rm{ELPS}}}

\newcommand{\Np}{N_{\mathrm{p}}}
\newcommand{\tsm}{t_{\mathrm{sm}}}
\newcommand{\tsc}{t_{\mathrm{sc}}}
\newcommand{\nel}{n_{\mathrm{e}}}

\theoremstyle{definition}

\newtheorem{rem}{Remark}

\begin{document}
	
\title{Efficient MPC for Emergency Evasive Maneuvers, Part II: Comparative Assessment for Hybrid Control}

\author{Leila~Gharavi,
	Bart~De~Schutter,~\IEEEmembership{Fellow, IEEE}
	and~Simone~Baldi,~\IEEEmembership{Senior Member, IEEE}
	\thanks{Leila~Gharavi and Bart~De~Schutter are with the Delft Center for Systems and Control, Delft University of Technology, 2628 CD Delft, The Netherlands (e-mails: \texttt{L.Gharavi@tudelft.nl}; \texttt{B.DeSchutter@tudelft.nl}).}
	\thanks{Simone~Baldi is with the School of Mathematics, Southeast University, Nanjing 21118, China (e-mail: \texttt{103009004@seu.edu.cn}).}}

% The paper headers
\markboth{Submitted to Journal, October~2024}%
{Shell \MakeLowercase{\textit{et al.}}: Bare Demo of IEEEtran.cls for IEEE Journals}

\maketitle

\begin{abstract}

Optimization-based approaches such as nonlinear Model Predictive Control (MPC) are promising approaches in safety-critical applications with nonlinear dynamics and uncertain environments such as automated driving systems. However, the computational complexity of the nonlinear MPC optimization problem coupled with the need for rapid response in emergency scenarios is the main bottleneck in realization of automation levels four and five for driving systems. In this paper, we construct hybrid formulations of the nonlinear MPC problem for vehicle control during emergency evasive maneuvers and assess their computational efficiency in terms of accuracy and solution time. To hybridize the MPC problem, we combine three hybrid approximations of the prediction model and four approximations of the nonlinear stability and tire saturation constraints and simulate the closed-loop behavior of the resulting controllers during five emergency maneuvers for different prediction horizons. Further, we compare the robustness of the controllers and their accuracy-time trade-off when the friction of the road is either unknown or has an offset error with respect to the prediction model. This robustness is investigated for different levels of friction uncertainty and with respect to the proximity to the vehicle handling limits. Our tests show that the hybridization of the MPC problem result in an efficient implementation of MPC for emergency evasive maneuvers, paving the way for implementation at high levels of automation.

\end{abstract}


\IEEEpeerreviewmaketitle

\section{Introduction} \label{sec:intro}
 
\IEEEPARstart{R}{eal-time} implementation of nonlinear MPC for high-speed safety-critical evasive maneuvers is an open research problem~\cite{Stano2023}. Two specific reasons contribute to this: high computation times for solving a NonLinear Program (NLP) compared to a linear or a Quadratic Program (QP), and possible convergence to local optima, which is highly sensitive to the initial guess provided to the NLP solver. 

Proactive vehicle control in emergency scenarios requires using the full control potential of the system, which means that some sub-optimal solution techniques for the NLP~\cite{Fu2022,Liu2022,Zhang2023} are not suitable to incorporate~\cite{Zhu2023}. To mitigate the slow convergence of NLP solvers, an upper bound is often imposed for the computation time as stopping criterion; this bound can be selected e.g., as a function of the complexity of the problem using prediction horizon, decision variable, etc. If the solver does not converge to a optimum before hitting this bound, the solution to the previous step is shifted and used~\cite{Chowdhri2021}. Nevertheless, if this occurs repeatedly and the controller does not converge to a solution for consecutive steps, this may result in a large degree of suboptimality or even infeasibility.

A popular approach for selecting the initial guess is using a warm-start strategy based on shifting the previous solution to tailor it for the current MPC optimization problem~\cite{Ardakani2019,Chowdhri2021,Guo2020}, which is suitable provided that the previous step converged to a good solution. This however is a restrictive condition, for which~\cite{Diehl2009} proposed using a tangential solution predictor instead of shifting, which is essentially based on using the concept of parametric sensitivity of the NLP for constructing new initial guesses. Nevertheless, warm start is a suitable strategy only if the solver converged to a ``good" solution in the previous step~\cite{Gros2020}. Other strategies to improve the initial guess include using the reference trajectory~\cite{Gros2020}, using the inverse static model of the system~\cite{Lawrynczuk2022}, or selecting the solution to a simpler approximation of the NLP e.g.,\ a~QP~\cite{Ghandriz2023}. Nevertheless, the mentioned approaches are not sufficient for real-time proactive control during emergency evasive maneuvers where a more extensive search in the decision space is required.

During emergency maneuvers, relying on one solution is restrictive: even with the improvements on the search direction and transformation, the search for the optimum would be limited within a neighborhood of the solution for the previous time-step. However, abrupt changes to the reference trajectory e.g.,\ due to sudden appearance of an obstacle on the road, require a more extended exploration of the search space to increase the likelihood of finding an acceptable optimum. In this sense,~\cite{Liu2017} uses a divide-and-conquer strategy in searching for starting regions based on the current state and then picks the first solution that satisfies an acceptable bound on the objective. While this method improves convergence to better optima, it still does not expand the search region in case of abrupt changes in the reference. In~\cite{Wurts2022}, multiple filtered random initial guesses are used to solve the NLP problem and in~\cite{Vaupel2020}, the NLP is solved offline and a dataset of ``good” initial guesses to be used in real time is learned, which could be an improvement upon relying on one solution without wasting additional computational effort on initial guesses with lower improvement value. However, this approach is only applicable in case there is sufficient and reliable data to learn such guesses, which is usually not currently available for vehicle control in hazardous scenarios.

While multi-start solution of the NLP improves the chances of converging to a suitable optimum to use the full control potential of the vehicle, it significantly increases the computation time, which is the main obstacle toward proactive control and real-time implementation of MPC during emergency scenarios. In this sense, hybridization of the nonlinear control optimization problem was proposed~\cite{Asarin2007,vanHuijgevoort2023} to balance the computational efficiency via the trade-off between accuracy and the convergence speed by using a hybrid systems formalism~\cite{Lunze2009} to express the prediction model and the nonlinear constraints. 

Hybrid MPC for vehicle tracking control has attracted attention as a potential solution to tackle the problem of computational efficiency~\cite{Besselmann2008,Zhang2015,Amir2017,Rokonuzzaman2021,Zhao2023}. Nevertheless, to the best of our knowledge, the capability of the hybridization approach in improving the computational efficiency of MPC has neither been assessed for highly-nonlinear prediction models, nor investigated during hazardous scenarios and aggressive evasive maneuvers. In this sense, such scenarios are particularly important since they require using the full control potential of the vehicle in its handling limits and the need for fast computation is critical in collision avoidance.

This work is the second part of the publication ``Efficient MPC for Emergency Evasive Maneuvers". In ``Part I: Hybridization of the Nonlinear Problem" we proposed an approach to approximate the prediction model and nonlinear physics-based constraints using a hybrid system formalism. In particular, we exploited the Max-Min-Plus-Scaling (MMPS) formulation to obtain a hybrid representation of the MPC optimization problem. 

The contributions of this paper are: 
\begin{enumerate}
	\item improving the computational efficiency of MPC,
	\item definition of a comparison benchmark,
	\item evaluation of several hybridization methods in the context of model predictive control, and 
	\item showcasing the impact and efficacy of hybridization in terms of control performance and computation speed.
\end{enumerate}
In this paper, we use the approximated prediction model and constraints to formulate and to solve the MPC problem as either a Mixed-Integer Linear Program (MILP) or a Mixed-Integer Quadratically-Constrained Program (MIQCP). We then investigate the trade-off between the accuracy and the computation speed of the resulting hybrid MPC controllers against their nonlinear counterparts. The computational performance of the hybrid and nonlinear controllers are assessed during five aggressive evasive maneuvers, representing abrupt changes in the reference trajectory due to a hazardous situation such as a sudden appearance of an obstacle on the road. Further, we investigate the tracking errors in the presence of uncertainty in the friction coefficient as an offset as well as a disturbance such as a significant decrease of friction due to the presence of water on a section of the road. 

This paper is organized as follows: the theoretical background such as the formulation of the nonlinear and hybrid MPC problems is explained in Section~\ref{sec:back}. To make this paper self-contained, we recall the hybridization approach in Part~I and its corresponding notation, to which the reader is referred. Section~\ref{sec:sim} explains different aspects of the comparison benchmark and assessment criteria e.g.,\ the choice of driving scenarios and the prediction horizons. The results of the simulations and the comparative assessment are discussed in Section~\ref{sec:res} followed by high-fidelity simulations in IPG CarMaker in Section~\ref{sec:ipg}. Finally, Section~\ref{sec:conc} presents the main results and draws an outlook for future research.

\section{Background}\label{sec:back}
\subsection{Model and Constraint Hybridization}
Consider a nonlinear discrete-time system
\begin{equation*}
	x (k+1) = F\left(x (k), u (k)\right),
\end{equation*}
where $x \in \mathbb{R}^n$ and $u \in \mathbb{R}^m$ represent the state and input vectors, respectively. We approximate each component $F_s$ of $F = \begin{bmatrix} F_1 & \dots & F_n \end{bmatrix}^T$ separately by an MMPS function $f_s$ with the Kripfganz form~\cite{Kripfganz1987} as
\begin{equation}
	\begin{split}
		f_s (x,u) = \max \left(\phi^+_s (x,u)\right) - \max \left(\phi^-_s (x,u)\right), \\ \forall s \in \{1, \dots, n\},
	\end{split}	
	\label{eq:mmpsdef}
\end{equation}
where the vectors $\phi^\eta_s: \mathbb{R}^{m+n} \to \mathbb{R}^{P^\eta}$ with $\eta \in \{+,-\}$ are affine functions of $x$ and $u$, also referred to as dynamic modes, and expressed via matrices
\begin{equation*}
	\begin{split}
		A^\eta_s \in \mathbb{R}^{P^\eta \times m}, \qquad
		B^\eta_s \in \mathbb{R}^{P^\eta \times n}, \qquad
		H^\eta_s \in \mathbb{R}^{P^\eta},\\
		\forall \eta \in \{+,-\}, \forall s \in \{1, \dots, n\},
	\end{split}
\end{equation*}
as
\begin{equation*}
	\phi^\eta_s (x,u) =  A^\eta_s x + B^\eta_s u + H^\eta_s.
\end{equation*}
The general form of $f$ is then given as
\begin{equation*}
	f(x,u) = \Psi^+ (x,u) - \Psi^- (x,u),
\end{equation*}
where $\Psi^+$ and $\Psi^-$ are vector-valued functions\footnote{In Part I, we used a scalar representation of the MMPS formulation to approximate each component of $F$ separately. In this paper, we use vector-valued representation to make the formulation of the MPC optimization problem more compact. However, without loss of generality, we consider $G$ to be a scalar function.} with
\begin{equation*}
	\Psi^\eta_s (x,u) = \max \left(\phi^\eta_s (x,u)\right), \quad \forall \eta \in \{+,-\}, \forall s \in \{1, \dots, n\}.
\end{equation*}
Note that in this notation, the max operator returns the largest component in the vector $\phi^\eta_s$.

For bounded $x$ and $u$, the physics-based constraints are in general nonlinear and non-convex and expressed via the normalized boundary function $G$ as
\begin{equation*}
	\mathcal{C} \coloneqq \{ (x,u) \in \mathbb{R}^{m+n} \; \vert \; 0 \leqslant G(x,u) \leqslant 1\},
\end{equation*}
where $\mathcal{C}$ is referred to as the feasible region. It should be noted that we normalize the constraint function to the interval $[0,1]$ to avoid numerical issues in the subsequent control optimization problems.
%\begin{rem}
%	Without loss of generality, we consider $G$ to be a scalar function. For more details, the reader is referred to Remark~1 in Part~I of this paper.
%\end{rem}
The region $\mathcal{C}$ is approximated either by a union of ellipsoids or by using the MMPS formalism, which corresponds to approximating the $\mathcal{C}$ by a union of convex polytopes.
In the MMPS approach, a similar formulation to the MMPS model approximation problem is used: we approximate $G$ by an MMPS function $\gl$ of the Kripfganz form (\ref{eq:mmpsdef}). The resulting feasible region $\Rl$ is then expressed as
\begin{equation}
	\Rl \coloneqq \{(x,u) \in \mathbb{R}^{m+n} \; | \; 0 \leqslant \gl(x,u) \leqslant 1\}.
	\label{eq:mmpscdef}
\end{equation}
via the boundary function
\begin{align}
	\gl (x,u) = \max \left(\gamma^+ (x,u)\right) - \max \left(\gamma^- (x,u)\right),
	\label{eq:mmpxgdef}
\end{align}
where $\gamma^\eta: \mathbb{R}^{m+n} \to \mathbb{R}^{R^\eta}$ are affine functions of $x$ and $u$ as
\begin{equation*}
	\gamma^\eta (x,u) =  C^\eta x + D^\eta u + I^\eta,
\end{equation*}
and
\begin{equation*}
	C^\eta \in \mathbb{R}^{R^\eta \times m}, \quad
	D^\eta \in \mathbb{R}^{R^\eta \times n}, \quad
	I^\eta \in \mathbb{R}^{R^\eta}, \qquad
	\forall \eta \in \{+,-\}.
\end{equation*}
The second way is to approximate the feasible region by a union of $n_{\rm{e}}$ ellipsoids as
\begin{align}
	\Rq \coloneqq \{(x,u) \in \mathbb{R}^{m+n} \; | \; 0 \leqslant \gq (x,u) \leqslant 1\},
	\label{eq:quniondef}
\end{align}
whose boundary can be expressed by
\begin{align}
	\gq (x,u) = \min \left(\omega (x,u) \right),
	\label{eq:ellipgdef}
\end{align}
where the min operator gives the smallest component in the vector $\omega$, and where
\begin{align}
	\omega_e (x,u) = \begin{pmatrix}
		x-x_{0,e} \\ u-u_{0,e}
	\end{pmatrix}^T Q_e \begin{pmatrix}
		x-x_{0,e} \\ u-u_{0,e}
	\end{pmatrix} - 1 , \\
	\qquad \forall e \in \{1,\dots, n_{\rm{e}}\},
	\label{eq:omegadef}
\end{align}
with $Q_e$ being a positive definite matrix and $(x_{0,e},u_{0,e})$ representing the center coordinates of the ellipsoid. Note that this representation includes rotated ellipsoids as well.

\subsection{MPC Optimization Problems}

The state and input vectors over the whole prediction horizon\footnote{In this study, for the sake of simplicity, the control horizon is assumed to be equal to the prediction horizon $\Np$.} $\Np$ are defined as
\[\tilde{x} (k+1) = \begin{bmatrix} \hat{x}^T(k +1|k) & \hat{x}^T (k+2 | k) & \dots & \hat{x}^T (k+\Np | k) \end{bmatrix}^T,\]
\[\tilde{u} (k) = \begin{bmatrix} u^T (k) & u^T (k+1) & \dots & u^T (k+\Np -1) \end{bmatrix}^T,\]
where $\hat{x} (k+i | k)$ represents the predicted state of the $(k+i)$-th time step based on the state measurement at the $k$-th time step. In addition and for brevity of expressions, we introduce the generalized form of the systems dynamics $F$ and inequality constraints $G$ over the prediction horizon as
\begin{equation*}
	\begin{split}
		\left[x (k+1) = F \left(x(k),u(k)\right)\right] 
		\iff 
		\left[\tilde{x}(k+1) = \tilde{F} \left(\tilde{x}(k),\tilde{u} (k)\right)\right],\\
		\left[0 \leqslant G \left(x(k),u(k)\right) \leqslant 1\right] 
		\iff 
		\left[0 \leqslant \tilde{G} \left(\tilde{x}(k),\tilde{u} (k)\right) \leqslant 1\right].
	\end{split}
\end{equation*}
Note that $\tilde{F}$ is the generalized counterpart of $F$ by extending the notation over the prediction horizon and not by recursive substitution. For the sake of brevity, $x(k)$ is not an explicit argument of $\tilde{F}$ but note that the dependence of $\tilde{F}$ on $x(k)$ is implied within the $\tilde{x}(k)$ argument.

Using the $\ell_1$-norm in defining the objective function in tracking $\tilde{x}_\text{ref}$, MPC requires solving the optimization problem 
\begin{align}
	\min_{\tilde{u} (k)} \quad & \Vert \Theta_x \left( \tilde{x} (k) - \tilde{x}_\text{ref} (k) \right) \Vert_1 + \Vert  \Theta_u \tilde{u} (k) \Vert_1,\\
	\text{s.t.} \quad &  \tilde{x}(k+1) = \tilde{F} \left(\tilde{x}(k),\tilde{u} (k)\right), \label{eq:nmpcm}\\
	& 0 \leqslant \tilde{G} \left(\tilde{x}(k),\tilde{u} (k)\right) \leqslant 1, \label{eq:nmpcc} 
\end{align}
with $\Theta_x \geqslant 0$ and $\Theta_u \geqslant 0$ being normalizing diagonal matrices with non-negative entries for the state tracking error and input signals, respectively. Note that the $\ell_1$-norm is selected to allow a mixed-integer linear description of the objective function.

The hybrid MPC problem can then be formulated as:
\begin{align}
	& \min_{\tilde{u} (k)} && \Theta_x \; \tilde{e}_x (k) \; + \; \Theta_u \;  \tilde{e}_u (k) \label{eq:hmpcj} \\
	& \text{s.t.} && -\tilde{e}_x (k) \leqslant \tilde{x} (k) - \tilde{x}_\text{ref} (k) \leqslant \tilde{e}_x (k), \label{eq:hmpce1}\\
	& &&-\tilde{e}_u (k) \leqslant \tilde{u} (k) \leqslant \tilde{e}_u (k),\label{eq:hmpce2}\\ 
	& && \tilde{x}(k+1) = \mathrm{vec} \left(\Psi^+ (k)\right) - \mathrm{vec} \left(\Psi^- (k)\right),\label{eq:hmpcm}\\
	& && \Psi^\eta_{ij} (k) = \max \left(\phi^\eta_i (k+j-1)\right),\label{eq:hmpcphi}\\
	& && \hspace{0.22\textwidth} \forall \eta \in \{+,-\}, \notag \\
	& && \hspace{0.22\textwidth} \forall i \in \{1, \dots, n\},\notag\\
	& && \hspace{0.22\textwidth} \forall j \in \{1, \dots, \Np\}, \notag
\end{align}
where (\ref{eq:hmpce1})--(\ref{eq:hmpce2}) are introduced to obtain a linear representation of the objective function by defining
\begin{align*}
	\tilde{e}_x (k) = \Vert \tilde{x} (k) - \tilde{x}_\text{ref} (k) \Vert_1, && \tilde{e}_u (k) = \Vert \tilde{u} (k) \Vert_1,
\end{align*}
and (\ref{eq:hmpcm})--(\ref{eq:hmpcphi}) are the hybridized model approximation to replace \eqref{eq:nmpcm}. The $\mathrm{vec}(\cdot)$ operator in (\ref{eq:hmpcm}) converts its matrix argument into a vector by stacking its components into one column vector. Then, constraint approximation can be hybridized by replacing \eqref{eq:nmpcc} by the MMPS constraints \eqref{eq:milp} for an MILP or the ellipsoidal constraints \eqref{eq:miqp} for an MIQCP formulation:
\begin{subequations}
	\begin{align}
		& \Lambda^\eta_j = \max \left(\gamma^\eta(k+j-1)\right),\label{eq:milp}\\
		& \hspace{0.22\textwidth} \forall \eta \in \{+,-\}, \notag\\
		& \hspace{0.22\textwidth} \forall j \in \{1, \dots, \Np\}, \notag\\
		& \Omega_j (k) = \min \left(\omega (k+j-1)\right),\label{eq:miqp}\\
		& \hspace{0.22\textwidth} \forall j \in \{1, \dots, \Np\}. \notag
	\end{align}
\end{subequations}

\begin{rem}
	The binary variables of the optimization problem are introduced via activating the local modes for the hybrid model and constraints (for more details, see Part~I).  Therefore, the corresponding MILP problem will have 
	\[\Np\left(R^+ + R^- + \sum_{s=1}^n (P_s^+ + P_s^-)\right)\]
	binary variables with $R^+$ and $R^-$ being the constraint-approximation counterparts of $P^+$ and $P^-$, and the MIQCP problem will have
	\[\Np\left(\nel + \sum_{s=1}^n (P_s^+ + P_s^-) \right)\]
	binary variables.
\end{rem}

\section{Comparison Benchmark}\label{sec:sim}% ~ ~ ~ ~ ~ ~ ~ ~ ~ ~ ~ ~ ~ ~ ~ ~ ~ ~ ~ ~ ~ ~ ~ ~ ~ ~ ~ ~ ~ ~ ~ ~ 

\subsection{Prediction Model and Physics-Based Constraints}

The nonlinear prediction model is a single-track vehicle with a Dugoff tire~\cite{Dugoff1970} model with varying friction as described in Appendix~\ref{app:model}, with system variables and parameters given in Tables~\ref{tab:vars} and \ref{tab:params}. 

In Part I of this paper, we hybridized the nonlinear model using different grid--generation methods via the MMPS formalism (\ref{eq:mmpsdef}) and obtained three hybrid approximations for the nonlinear model labeled by their corresponding grid types as R, S, and T models. The nonlinear physics-based constraints due to the tire force saturation and vehicle stability were hybridized as well via approximating the feasible region by a union of ellipsoids and by a union of polytopes (using the MMPS formalism) via boundary-based and region-based approximations. There, we obtained four approximations labeled by their approach (R and B) and by the shape of the subregions (MP for MMPS or EL for ellipsoidal) as RMP, BMP, REL, and BEL. Table~\ref{tab:abbrev} summarizes the abbreviations used in this paper for different hybrid models, constraints, and their corresponding controllers. For more details of the boundary-based and region-based approximations and their errors, or the number of introduced binary variables by each approach, the reader is referred to Part I of this paper.
\begin{table}[htbp]
\caption{Abbreviations for hybrid models and controllers}
\begin{center}
	\begin{tabular}{c|c|c}
		\toprule
		\multicolumn{3}{c}{\emph{General}}\\
		\midrule
		\textbf{MILP} & \multicolumn{2}{c}{Mixed-Integer Linear Program}\\
		\textbf{MIQCP} & \multicolumn{2}{c}{Mixed-Integer Quadratically-Constrained Program}\\
		\textbf{MMPS} & \multicolumn{2}{c}{Max-Min-Plus-Scaling}\\
		\textbf{MPC} & \multicolumn{2}{c}{Model Predictive Control}\\
		\textbf{NLP} & \multicolumn{2}{c}{NonLinear Program}\\
		\textbf{QP} & \multicolumn{2}{c}{Quadratic Program}\\	
		\midrule[0.8pt]
		\multicolumn{3}{c}{\emph{Hybrid Models}}\\
		\midrule
		\multicolumn{2}{c|}{Approximation Grid Type} & Abbreviation\\
		\hline
		\multicolumn{2}{c|}{Domain-based random} & \textbf{R}\\
		\multicolumn{2}{c|}{Trajectory-based steady-state initiated} & \textbf{S}\\
		\multicolumn{2}{c|}{Trajectory-based randomly initiated} & \textbf{T}\\
		\midrule[0.8pt]
		\multicolumn{3}{c}{\emph{Hybrid Constraints}}\\
		\midrule
		& \multicolumn{2}{c}{Formulation}\\\cline{2-3}
		& Ellipsoidal (EL) & MMPS (MP)\\
		\hline
		Region-based (R) &  \textbf{REL} & \textbf{RMP} \\
		Boundary-based (B) & \textbf{BEL} & \textbf{BMP} \\
		\midrule[0.8pt]
		\multicolumn{3}{c}{\emph{Hybrid MPC Controllers}}\\
		\midrule
		\multicolumn{3}{c}{[Model abbreviation] -- [Constraint abbreviation]}\\
		\multicolumn{3}{c}{Example: \textbf{R} model + \textbf{BMP} constraint $\to$ \textbf{R--BMP} controller}\\
		\bottomrule
	\end{tabular}
	\label{tab:abbrev}
\end{center}
\end{table}

\subsection{Control Parameters}

Given the application, the \emph{control sampling time} $\tsc$ is often restricted by the capabilities of the control hardware such as the maximum operation frequency. In addition, the \emph{model sampling time} $\tsm$ is either known a priori for a discrete-time system, or obtained for a continuous-time system with respect to its natural frequency and dynamic behavior. Therefore, we assume these two parameters to be fixed during all the simulations as known system parameters $\tsm = 0.01\rm{s}$ and $\tsc = 0.05\rm{s}$. 

In the path tracking MPC literature~\cite{Stano2023}, the time span of the prediction often covers below 1.5s ahead, for control sampling times shorter than 0.1s. Based on our selected control sampling time $\tsc = 0.05\rm{s}$, we therefore test different time spans of the prediction in the range 0.25s to 1.50s, corresponding to $\Np \in \{5, 10, \dots, 30\}$. This is further explained in Section~\ref{sec:scenarios}.

\subsection{MPC Controllers}

In this benchmark, we consider two nonlinear MPC controllers with the nonlinear prediction model and the physics-based constraints. The first one solves the NLP using the warm start only (i.e.\ with the shifted solution of the previous time step) labeled as NL--1, and the second one referred to as NL--5 solves the problem for five different initial guesses and selects the best solution it has found. The initial guesses for NL--5 are as follows: one warm start as in NL--1, one random point within the domain, one point at the lower bound, one at the upper bound, and one in the geometric center. The solution for each control time step is fed to the system via a receding-horizon strategy: while the control optimization problem finds the optimal input signal during the next $N_{\rm{p}}$ control time steps, we only use the solution for the first step.

The computational performance of the two NLP controllers is compared against MPC controllers based on combinations of three hybrid models (R, S, and T) with four constraint approximations: two polytopic (RMP and BMP) and two ellipsoidal (REL and BEL). In total we have six MILP and six MIQCP controllers to compare against their NLP counterparts. The hybrid MPC controllers are labeled by combining the abbreviations for their model and constraints, separated by a dash (--) symbol, as described in Table~\ref{tab:abbrev}. Further, MILP and MIQCP controllers can be constructed using our published hybridization toolbox~\cite{HybridCode}.

\subsection{Reference Trajectory}\begin{figure}[t]
\begin{center}
%\begin{subfigure}[t]{\textwidth}
%\begin{tikzpicture}
%\begin{axis}[width=0.23\textwidth,height=0.23\textwidth,	
%	xlabel={$v_x$ (m/s)},ylabel={$v_y$ (m/s)},
%	xmin=17,xmax=45,ymin=-3,ymax=1,
%	xmajorgrids=true, xminorgrids=true,ymajorgrids=true,yminorgrids=true,
%	grid style=dashed]	
%	\addplot[color=black,ultra thick,decoration={markings,mark=between positions 0.5 and 0.6 step 5em with {\arrow [scale=1.0]{to}}},postaction=decorate] table [x=VX,y=VY]{Files/REF_MAN1.dat};	
%	\addplot[color=black,ultra thick,decoration={markings,mark=between positions 0.5 and 0.6 step 5em with {\arrow [scale=1.0]{to}}},postaction=decorate] table [x=VX,y=VY]{Files/REF_MAN2.dat};	
%	\addplot[color=black,ultra thick,decoration={markings,mark=between positions 0.5 and 0.6 step 5em with {\arrow [scale=1.0]{to}}},postaction=decorate] table [x=VX,y=VY]{Files/REF_MAN3.dat};	
%	\addplot[color=black,ultra thick,decoration={markings,mark=between positions 0.5 and 0.6 step 5em with {\arrow [scale=1.0]{to}}},postaction=decorate] table [x=VX,y=VY]{Files/REF_MAN4.dat};	
%	\addplot[color=black,ultra thick,decoration={markings,mark=between positions 0.5 and 0.6 step 5em with {\arrow [scale=1.0]{to}}},postaction=decorate] table [x=VX,y=VY]{Files/REF_MAN5.dat};	
%\end{axis}
%\end{tikzpicture}
\begin{subfigure}[t]{\textwidth}
\begin{tikzpicture}
\begin{axis}[width=0.24\textwidth,height=0.24\textwidth,	
	xlabel={$a_x$ (m/s$^2$)},ylabel={$a_y$ (m/s$^2$)},
	xmin=-10,xmax=10,ymin=-10,ymax=10,
	xmajorgrids=true, xminorgrids=true,ymajorgrids=true,yminorgrids=true,
	grid style=dashed]		
	\filldraw[color=red,fill=red,fill opacity=0.1,thick,dashed](0,0) circle (8);
	\filldraw[color=gray,fill=green,fill opacity=0.1,thick,dashed](0,0) circle (5);
	\addplot[color=cyan,ultra thick,decoration={markings,mark=between positions 0.1 and 0.9 step 2em with {\arrow [scale=1.0]{to}}},postaction=decorate] table [x=AX,y=AY]{Files/REF_MAN1.dat};
	\addplot[color=magenta,ultra thick,decoration={markings,mark=between positions 0.1 and 0.9 step 2em with {\arrow [scale=1.0]{to}}},postaction=decorate] table [x=AX,y=AY]{Files/REF_MAN2.dat};
	\addplot[color=blue,ultra thick,decoration={markings,mark=between positions 0.1 and 0.9 step 2em with {\arrow [scale=1.0]{to}}},postaction=decorate] table [x=AX,y=AY]{Files/REF_MAN3.dat};
	\addplot[color=orange,ultra thick,decoration={markings,mark=between positions 0.1 and 0.9 step 2em with {\arrow [scale=1.0]{to}}},postaction=decorate] table [x=AX,y=AY]{Files/REF_MAN4.dat};
	\addplot[color=violet,ultra thick,decoration={markings,mark=between positions 0.1 and 0.9 step 2em with {\arrow [scale=1.0]{to}}},postaction=decorate] table [x=AX,y=AY]{Files/REF_MAN5.dat};
\end{axis}
\end{tikzpicture}
\begin{tikzpicture}
\begin{axis}[width=0.24\textwidth,height=0.24\textwidth,	
	xlabel={$\beta$ (rad)},ylabel={$r$ (rad/s)},
	xmin=-0.1,xmax=0.1,ymin=-0.8,ymax = 0.8,xtick={-0.06,0,0.06},scaled ticks=false,tick label style={/pgf/number format/fixed},
	xmajorgrids=true, xminorgrids=true,ymajorgrids=true,yminorgrids=true,
	grid style=dashed]		
	\filldraw[color=gray,fill=green,fill opacity=0.1,thick,dashed](-0.08,-0.32) rectangle (0.08,0.32);
	\addplot[color=cyan,ultra thick,decoration={markings,mark=between positions 0.4 and 0.6 step 2em with {\arrow [scale=1.0]{to}}},postaction=decorate] table [x=BETA,y=R]{Files/REF_MAN1.dat};
	\addplot[color=magenta,ultra thick,decoration={markings,mark=between positions 0.4 and 0.6 step 2em with {\arrow [scale=1.0]{to}}},postaction=decorate] table [x=BETA,y=R]{Files/REF_MAN2.dat};
	\addplot[color=blue,ultra thick,decoration={markings,mark=between positions 0.4 and 0.6 step 2em with {\arrow [scale=1.0]{to}}},postaction=decorate] table [x=BETA,y=R]{Files/REF_MAN3.dat};
	\addplot[color=orange,ultra thick,decoration={markings,mark=between positions 0.4 and 0.6 step 2em with {\arrow [scale=1.0]{to}}},postaction=decorate] table [x=BETA,y=R]{Files/REF_MAN4.dat};
	\addplot[color=violet,ultra thick,decoration={markings,mark=between positions 0.4 and 0.6 step 2em with {\arrow [scale=1.0]{to}}},postaction=decorate] table [x=BETA,y=R]{Files/REF_MAN5.dat};
\end{axis}
\end{tikzpicture}
\end{subfigure}
\subcaption{Stability constraints (g--g and $\beta$-$r$)}\label{fig:manstab}
\vspace*{5pt}
\begin{subfigure}[b]{\textwidth}\hspace*{10pt}
\begin{tikzpicture}
\begin{axis}[width=0.31\textwidth,height=0.31\textwidth,	
	xlabel={$F_x$ (kN)},ylabel={$F_y$ (kN)},
	xmin=-8.1,xmax=8.1,ymin=-8.1,ymax=8.1,
	xmajorgrids=true, xminorgrids=true,ymajorgrids=true,yminorgrids=true,
	grid style=dashed,legend pos=outer north east]		
	\filldraw[color=gray,fill=green,fill opacity=0.1,thick,dashed](0,0) circle (8);
	\addplot[color=cyan,dashed,ultra thick,decoration={markings,mark=between positions 0.2 and 0.9 step 3em with {\arrow [scale=1.0]{to}}},postaction=decorate] table [x=FXF,y=FYF]{Files/REF_MAN1.dat};
	\addplot[color=cyan,ultra thick,decoration={markings,mark=between positions 0.1 and 0.9 step 3em with {\arrow [scale=1.0]{to}}},postaction=decorate] table [x=FXR,y=FYR]{Files/REF_MAN1.dat};
	\addplot[color=magenta,dashed,ultra thick,decoration={markings,mark=between positions 0.2 and 0.9 step 3em with {\arrow [scale=1.0]{to}}},postaction=decorate] table [x=FXF,y=FYF]{Files/REF_MAN2.dat};
	\addplot[color=magenta,ultra thick,decoration={markings,mark=between positions 0.1 and 0.9 step 3em with {\arrow [scale=1.0]{to}}},postaction=decorate] table [x=FXR,y=FYR]{Files/REF_MAN2.dat};
	\addplot[color=blue,dashed,ultra thick,decoration={markings,mark=between positions 0.2 and 0.9 step 3em with {\arrow [scale=1.0]{to}}},postaction=decorate] table [x=FXF,y=FYF]{Files/REF_MAN3.dat};
	\addplot[color=blue,ultra thick,decoration={markings,mark=between positions 0.1 and 0.9 step 3em with {\arrow [scale=1.0]{to}}},postaction=decorate] table [x=FXR,y=FYR]{Files/REF_MAN3.dat};
	\addplot[color=orange,dashed,ultra thick,decoration={markings,mark=between positions 0.2 and 0.9 step 3em with {\arrow [scale=1.0]{to}}},postaction=decorate] table [x=FXF,y=FYF]{Files/REF_MAN4.dat};
	\addplot[color=orange,ultra thick,decoration={markings,mark=between positions 0.1 and 0.9 step 3em with {\arrow [scale=1.0]{to}}},postaction=decorate] table [x=FXR,y=FYR]{Files/REF_MAN4.dat};
	\addplot[color=violet,dashed,ultra thick,decoration={markings,mark=between positions 0.2 and 0.9 step 3em with {\arrow [scale=1.0]{to}}},postaction=decorate] table [x=FXF,y=FYF]{Files/REF_MAN5.dat};
	\addplot[color=violet,ultra thick,decoration={markings,mark=between positions 0.1 and 0.9 step 3em with {\arrow [scale=1.0]{to}}},postaction=decorate] table [x=FXR,y=FYR]{Files/REF_MAN5.dat};
	\legend{Man.\ 1 Front, Man.\ 1 Rear, Man.\ 2 Front, Man.\ 2 Rear, Man.\ 3 Front, Man.\ 3 Rear, Man.\ 4 Front, Man.\ 4 Rear, Man.\ 5 Front, Man.\ 5 Rear}
\end{axis}
\end{tikzpicture}
\end{subfigure}
\subcaption{Front/rear tire saturation constraint (Kamm circle)}\label{fig:mankamm}
\end{center}
\caption{Selected maneuvers for the benchmark, represented in terms of the constraints. The green zone in the g-g diagram represents the safe region and the red one corresponds to the aggressive yet acceptable acceleration range.}\label{fig:mans}
\end{figure}
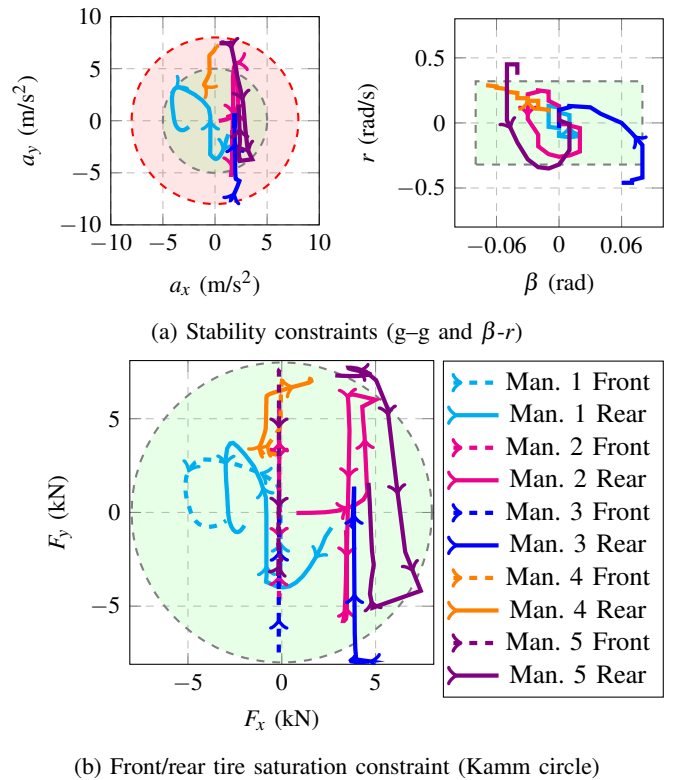

We compare the computational performance of the nonlinear and hybrid MPC controllers during five maneuvers of two seconds as reference trajectories. These maneuvers are selected to represent aggressive evasive maneuvers with different longitudinal velocities as explained in Table~\ref{tab:man}. Note that for a more intuitive representation of distance to the handling limits and/or tire saturation, we indicate the normalized distance of the reference trajectory in the g--g diagram (\ref{eq:gg}) and Kamm circles (\ref{eq:kamm}) in their respective column where 1 represents the boundary of the circles, i.e. the constraints. The column $\beta$-$r$ represents the normalized location of the stability envelope often used in the literature as \mbox{$| r | \leqslant \mu g v_x$} under the assumption of steady-state cornering conditions as a typical yaw rate constraint~\cite{Stano2023} and \mbox{$| \beta | \leqslant 5$deg}~\cite{Chowdhri2021}. Note that the steady-state cornering conditions do not hold in extreme maneuvers. Nevertheless, we provide the trajectory plots on the $\beta$-$r$ envelope to provide more insight in the cornering behavior of the vehicle during the reference maneuvers. Moreover, the normalized values represent the distance to the boundary with values between 0 and 1, where 1 indicates the position of the boundary itself. Figure~\ref{fig:mans} shows the schematic view of the five reference maneuvers in terms of these constraints. The two-seconds simulation time is selected to represent the recovery window for the controller in hazardous scenarios in case of an abrupt change in the reference trajectory.
\begin{table}[htb]		
	\caption{Selected maneuvers as reference trajectories for the benchmark. The $v_x$ column represents the average longitudinal velocity in km/h.}
	\label{tab:man}
	\begin{center}
		\begin{tabular}{c | c | c  c  c  c}
			\toprule
			\textbf{No.} & \textbf{Maneuver} & $\bold{v_x}$ & \textbf{g--g} & $\bold{\beta}$-$\bold{r}$ & \textbf{Kamm} \\
			\midrule
			1 & Safe lane change & 130  & 0--0.5 & 0--0.4 & 0--0.7 \\				
			2 & Aggressive lane change & 128 & 0--0.6 & 0--0.9 & 0--0.9 \\
			3 & Drift cornering & 73 & 0.2--0.8 & 0--0.9 & 0--1.0 \\
			4 & High-speed cornering &  154  & 0.3--0.7 & 0.3--1.1 & 0.4--0.9 \\
			5 & Low-speed cornering & 75  & 0.2--0.8 & 0.2--0.8 & 0--1.0 \\
			\bottomrule
		\end{tabular}
	\end{center}
\end{table}
\vspace{-0.5cm}
\subsection{Driving Scenarios}\label{sec:scenarios}

We compare the computational performance of the nonlinear and hybrid MPC controllers in the following four driving scenarios:
\begin{itemize}
	\item \textbf{Ideal Case:} The nonlinear prediction model is selected as the real system. The computational performance of the hybrid controllers is evaluated over $\Np \in \{5,10,\dots,30\}$.
	\item \textbf{Friction Offset:} We use a different tire-road friction in the real system as $\mu \in \{0.70, 0.75, \dots,1.00\}$ and compare the computational performance of the controllers for a selected $\Np$ value over this range of friction offsets. 
	\item \textbf{Friction Disturbance:} We assume the road friction for the second quarter of the maneuver to be very low, representing a disruption such as a slippery road surface and we compare the computational efficiencies in the same fashion as the friction offset case.
	\item \textbf{Handling Limits:} We investigate the computational performance of the controller for a fixed friction offset during the lane change maneuver (no.\ 1 in Table~\ref{tab:man}). We vary the input steering angle during the maneuver to simulate different levels of aggressive steering and assess the closed-loop performance in terms of the closeness to handling limits, i.e.,\ the boundary of the g--g diagram with the acceleration magnitude between 0.5$g$ up to the $\mu g$ limit.
\end{itemize}

Note that $\delta$ is the steering angle on the road and not the steering wheel angle, and since the scenarios are high speed where compared to an urban scenario, a smaller steering angle can lead to a higher lateral acceleration. For instance, the steering angle bound is approximately 10 degrees in~\cite{Chowdhri2021} and 20 degrees in \cite{Brown2017} or the extreme maneuvers in~\cite{Subosits2021}. For drifting, to the best of our knowledge, the maximum bound is 38 degrees in~\cite{Goh2020} where the velocity is 54 km/h, while we investigate 75 up to 154 km/h maneuvers. Therefore, we believe 30 degrees is a suitable upper bound for the steering angle for extreme maneuvers in highway scenarios.

\subsection{Solver Selection}

For a fair comparison in terms of computation time, we select the most efficient known solvers within the academic community for the MILP/MIQCP and NLP problems. 

The MILP/MIQCP and NLP problems are solved by GUROBI~\cite{GUROBI} and TOMLAB/KNITRO~\cite{TOMLAB} optimizers, respectively, using \textsc{Matlab} as interface and overall computation environment. To further improve the solution time for the NLP problems, we provided the objective and constraint functions via MEX files (instead of m-files in \textsc{Matlab}), which in our experiments reduced the computation time for the NLP problems by around 50\% for all the cases.

The simulations were all run on a PC with a 4-core(s) Intel Xeon 3.60 GHz CPU and 8 GB RAM on Windows 10 64-bit and in a \textsc{Matlab} R2020b environment.

% MANEUVER 2 : AGRESSIVE LANE CHANGE

\begin{figure*}[bth]
\begin{center}
\begin{subfigure}[b]{\linewidth}
	\begin{tikzpicture}
		\begin{semilogyaxis}[width=0.3\textwidth,height=0.23\textwidth,	
			xlabel={$\Np$},ylabel={Mean Rel. Err. (\%)},
			xmin=4, xmax=31,ymin=0, ymax=16,ytick={2,5,10,15},scaled ticks=false,tick label style={/pgf/number format/fixed},log ticks with fixed point,xmajorgrids=true, xminorgrids=true,ymajorgrids=true,yminorgrids=true,grid style=dashed]		
			\addplot[smooth,color=NL1,thick,mark=*] table [x=NP,y=NL1_MEAN]{Files/IDEAL_MAN2.dat};	
			\addplot[smooth,color=NL5,thick,mark=*] table [x=NP,y=NL5_MEAN]{Files/IDEAL_MAN2.dat};	
			\addplot[smooth,color=RRMP,thick,mark=*] table [x=NP,y=RRMP_MEAN]{Files/IDEAL_MAN2.dat};	
			\addplot[smooth,color=RBMP,thick,mark=*] table [x=NP,y=RBMP_MEAN]{Files/IDEAL_MAN2.dat};
			\addplot[smooth,color=SRMP,thick,mark=*] table [x=NP,y=SRMP_MEAN]{Files/IDEAL_MAN2.dat};	
			\addplot[smooth,color=SBMP,thick,mark=*] table [x=NP,y=SBMP_MEAN]{Files/IDEAL_MAN2.dat};	
			\addplot[smooth,color=TRMP,thick,mark=*] table [x=NP,y=TRMP_MEAN]{Files/IDEAL_MAN2.dat};	
			\addplot[smooth,color=TBMP,thick,mark=*] table [x=NP,y=TBMP_MEAN]{Files/IDEAL_MAN2.dat};	
		\end{semilogyaxis}
	\end{tikzpicture}
	\begin{tikzpicture}
		\begin{semilogyaxis}[width=0.3\textwidth,height=0.23\textwidth,	
			xlabel={$\Np$},ylabel={Max Rel. Err. (\%)},
			xmin=4, xmax=31,ymin=0, ymax=60,ytick={2,5,10,25,50},scaled ticks=false,tick label style={/pgf/number format/fixed},log ticks with fixed point,xmajorgrids=true, xminorgrids=true,ymajorgrids=true,yminorgrids=true,grid style=dashed]		
			\addplot[smooth,color=NL1,thick,mark=*] table [x=NP,y=NL1_MAX]{Files/IDEAL_MAN2.dat};	
			\addplot[smooth,color=NL5,thick,mark=*] table [x=NP,y=NL5_MAX]{Files/IDEAL_MAN2.dat};	
			\addplot[smooth,color=RRMP,thick,mark=*] table [x=NP,y=RRMP_MAX]{Files/IDEAL_MAN2.dat};	
			\addplot[smooth,color=RBMP,thick,mark=*] table [x=NP,y=RBMP_MAX]{Files/IDEAL_MAN2.dat};	
			\addplot[smooth,color=SRMP,thick,mark=*] table [x=NP,y=SRMP_MAX]{Files/IDEAL_MAN2.dat};	
			\addplot[smooth,color=SBMP,thick,mark=*] table [x=NP,y=SBMP_MAX]{Files/IDEAL_MAN2.dat};
			\addplot[smooth,color=TRMP,thick,mark=*] table [x=NP,y=TRMP_MAX]{Files/IDEAL_MAN2.dat};	
			\addplot[smooth,color=TBMP,thick,mark=*] table [x=NP,y=TBMP_MAX]{Files/IDEAL_MAN2.dat};									
		\end{semilogyaxis}
	\end{tikzpicture}
	\begin{tikzpicture}
		\begin{semilogyaxis}[width=0.3\textwidth,height=0.23\textwidth,	
			xlabel={$\Np$},ylabel={Comp. Time (s)},
			%xmin=1e-1, xmax=1e2,ymin=0, ymax=0.060,
			xmajorgrids=true, xminorgrids=true,ymajorgrids=true,yminorgrids=true,
			grid style=dashed,legend pos=outer north east]		
			\addplot[smooth,color=NL1,thick,mark=*] table [x=NP,y=NL1_TIME]{Files/IDEAL_MAN2.dat};	
			\addplot[smooth,color=NL5,thick,mark=*] table [x=NP,y=NL5_TIME]{Files/IDEAL_MAN2.dat};	
			\addplot[smooth,color=RRMP,thick,mark=*] table [x=NP,y=RRMP_TIME]{Files/IDEAL_MAN2.dat};	
			\addplot[smooth,color=RBMP,thick,mark=*] table [x=NP,y=RBMP_TIME]{Files/IDEAL_MAN2.dat};
			\addplot[smooth,color=SRMP,thick,mark=*] table [x=NP,y=SRMP_TIME]{Files/IDEAL_MAN2.dat};	
			\addplot[smooth,color=SBMP,thick,mark=*] table [x=NP,y=SBMP_TIME]{Files/IDEAL_MAN2.dat};	
			\addplot[smooth,color=TRMP,thick,mark=*] table [x=NP,y=TRMP_TIME]{Files/IDEAL_MAN2.dat};	
			\addplot[smooth,color=TBMP,thick,mark=*] table [x=NP,y=TBMP_TIME]{Files/IDEAL_MAN2.dat};
			\legend{NL--1, NL--5, R--RMP, R--BMP, S--RMP, S--BMP, T--RMP,T--BMP}
		\end{semilogyaxis}
	\end{tikzpicture}
	\subcaption{MILP vs.\ NLP controllers}
	\label{fig:ideal2milp}
\end{subfigure}		
\begin{subfigure}[b]{\linewidth}
	\begin{tikzpicture}
		\begin{semilogyaxis}[width=0.3\textwidth,height=0.23\textwidth,	
			xlabel={$\Np$},ylabel={Mean Rel. Err. (\%)},
			xmin=4, xmax=31,ymin=0, ymax=16,ytick={2,5,10,15},scaled ticks=false,tick label style={/pgf/number format/fixed},log ticks with fixed point,xmajorgrids=true, xminorgrids=true,ymajorgrids=true,yminorgrids=true,grid style=dashed]		
			\addplot[smooth,color=NL1,thick,mark=*] table [x=NP,y=NL1_MEAN]{Files/IDEAL_MAN2.dat};	
			\addplot[smooth,color=NL5,thick,mark=*] table [x=NP,y=NL5_MEAN]{Files/IDEAL_MAN2.dat};	
			\addplot[smooth,color=RRMP,thick,mark=*] table [x=NP,y=RREL_MEAN]{Files/IDEAL_MAN2.dat};	
			\addplot[smooth,color=RBMP,thick,mark=*] table [x=NP,y=RBEL_MEAN]{Files/IDEAL_MAN2.dat};
			\addplot[smooth,color=SRMP,thick,mark=*] table [x=NP,y=SREL_MEAN]{Files/IDEAL_MAN2.dat};	
			\addplot[smooth,color=SBMP,thick,mark=*] table [x=NP,y=SBEL_MEAN]{Files/IDEAL_MAN2.dat};	
			\addplot[smooth,color=TRMP,thick,mark=*] table [x=NP,y=TREL_MEAN]{Files/IDEAL_MAN2.dat};	
			\addplot[smooth,color=TBMP,thick,mark=*] table [x=NP,y=TBEL_MEAN]{Files/IDEAL_MAN2.dat};	
		\end{semilogyaxis}
	\end{tikzpicture}
	\begin{tikzpicture}
		\begin{semilogyaxis}[width=0.3\textwidth,height=0.23\textwidth,	
			xlabel={$\Np$},ylabel={Max Rel. Err. (\%)},
			xmin=4, xmax=31,ymin=0, ymax=60,ytick={2,5,10,25,50},scaled ticks=false,tick label style={/pgf/number format/fixed},log ticks with fixed point,xmajorgrids=true, xminorgrids=true,ymajorgrids=true,yminorgrids=true,grid style=dashed]		
			\addplot[smooth,color=NL1,thick,mark=*] table [x=NP,y=NL1_MAX]{Files/IDEAL_MAN2.dat};	
			\addplot[smooth,color=NL5,thick,mark=*] table [x=NP,y=NL5_MAX]{Files/IDEAL_MAN2.dat};	
			\addplot[smooth,color=RRMP,thick,mark=*] table [x=NP,y=RREL_MAX]{Files/IDEAL_MAN2.dat};	
			\addplot[smooth,color=RBMP,thick,mark=*] table [x=NP,y=RBEL_MAX]{Files/IDEAL_MAN2.dat};	
			\addplot[smooth,color=SRMP,thick,mark=*] table [x=NP,y=SREL_MAX]{Files/IDEAL_MAN2.dat};	
			\addplot[smooth,color=SBMP,thick,mark=*] table [x=NP,y=SBEL_MAX]{Files/IDEAL_MAN2.dat};
			\addplot[smooth,color=TRMP,thick,mark=*] table [x=NP,y=TREL_MAX]{Files/IDEAL_MAN2.dat};	
			\addplot[smooth,color=TBMP,thick,mark=*] table [x=NP,y=TBEL_MAX]{Files/IDEAL_MAN2.dat};									
		\end{semilogyaxis}
	\end{tikzpicture}
	\begin{tikzpicture}
		\begin{semilogyaxis}[width=0.3\textwidth,height=0.23\textwidth,	
			xlabel={$\Np$},ylabel={Comp. Time (s)},
			%xmin=1e-1, xmax=1e2,ymin=0, ymax=0.060,
			xmajorgrids=true, xminorgrids=true,ymajorgrids=true,yminorgrids=true,grid style=dashed,legend pos=outer north east]		
			\addplot[smooth,color=NL1,thick,mark=*] table [x=NP,y=NL1_TIME]{Files/IDEAL_MAN2.dat};	
			\addplot[smooth,color=NL5,thick,mark=*] table [x=NP,y=NL5_TIME]{Files/IDEAL_MAN2.dat};	
			\addplot[smooth,color=RRMP,thick,mark=*] table [x=NP,y=RREL_TIME]{Files/IDEAL_MAN2.dat};	
			\addplot[smooth,color=RBMP,thick,mark=*] table [x=NP,y=RBEL_TIME]{Files/IDEAL_MAN2.dat};
			\addplot[smooth,color=SRMP,thick,mark=*] table [x=NP,y=SREL_TIME]{Files/IDEAL_MAN2.dat};	
			\addplot[smooth,color=SBMP,thick,mark=*] table [x=NP,y=SBEL_TIME]{Files/IDEAL_MAN2.dat};	
			\addplot[smooth,color=TRMP,thick,mark=*] table [x=NP,y=TREL_TIME]{Files/IDEAL_MAN2.dat};	
			\addplot[smooth,color=TBMP,thick,mark=*] table [x=NP,y=TBEL_TIME]{Files/IDEAL_MAN2.dat};
			\legend{NL--1, NL--5, R--REL, R--BEL, S--REL, S--BEL, T--REL,T--BEL}
		\end{semilogyaxis}
	\end{tikzpicture}
	\subcaption{MIQCP vs.\ NLP controllers}
	\label{fig:ideal2miqcp}
\end{subfigure}
\caption{Computational performance of the nonlinear and hybrid MPC controllers during the aggressive lane change maneuver (maneuver 2 in Table~\ref{tab:man}) in the ideal scenario.}
\label{fig:ideal2}	
\end{center}
\end{figure*}
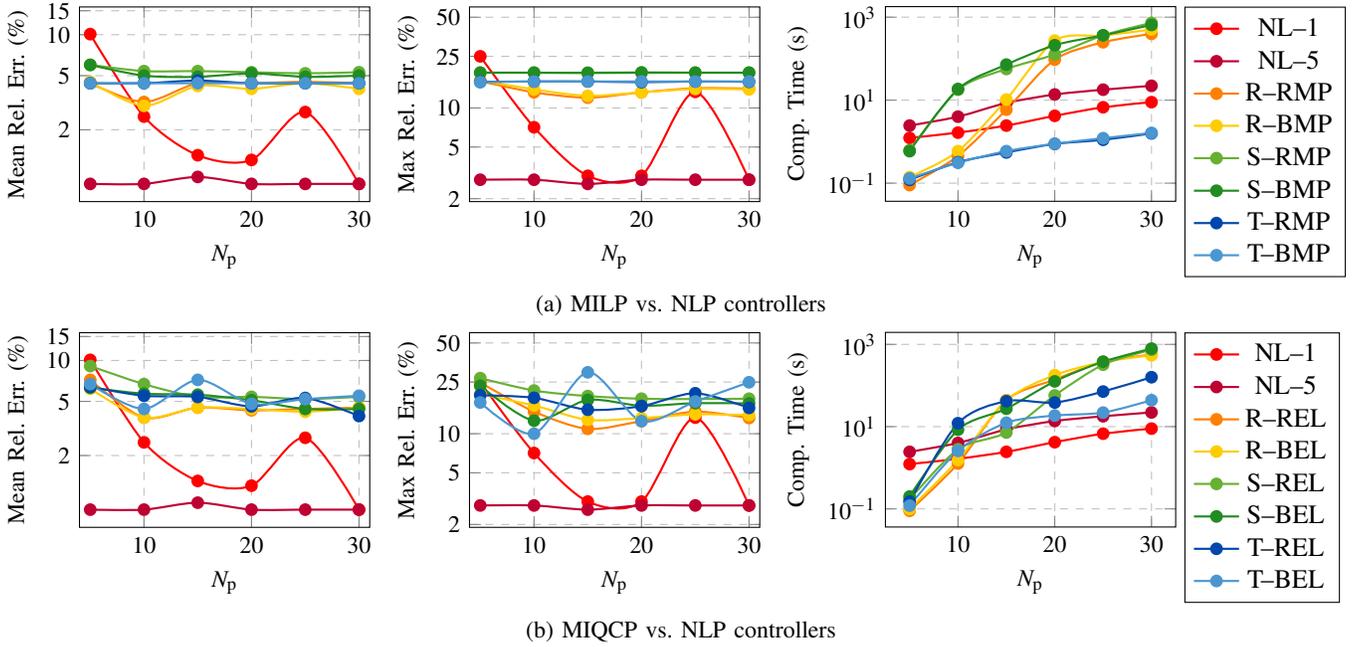

\section{Simulation Results}\label{sec:res}% ~ ~ ~ ~ ~ ~ ~ ~ ~ ~ ~ ~ ~ ~ ~ ~ ~ ~ ~ ~ ~ ~ ~ ~

Using the benchmark described in Section~\ref{sec:sim}, we compare the computational performance of the hybrid and nonlinear MPC controllers as follows: we first compare these controllers in the ideal case for different $\Np$ values and then we select the most promising hybrid controllers and compare their robustness to friction uncertainty and their performance close to the handling limits in the next subsections.

\subsection{Ideal Case}\label{sec:subideal}
The computational performance of the nonlinear and hybrid MPC controllers is shown in Figures~\ref{fig:ideal2} and \ref{fig:ideal345} in terms of their average and maximum tracking errors and computation time per control sampling time. 

\subsubsection{Aggressive Lane Change Maneuver}

First, we start by comparing the hybrid and nonlinear MPC controllers during the aggressive lane change maneuver in Fig.~\ref{fig:mans}.

In this maneuver, all MILP controllers, as well as NL--5, perform equally well with different $N_{\rm{p}}$ values. However, NL–1 shows oscillatory tracking accuracy across the $N_{\rm{p}}$ axis. This happens because extending the prediction horizon, even with the same prediction model as the ground truth, makes the search for an optimal solution more difficult, which reduces the chance of reaching an “acceptable” solution when only using the warm-start strategy, especially when performing the aggressive lane change maneuver shown in Fig.~\ref{fig:mans}.

Extending the prediction horizon not only enlarges the optimization problem's search space but also causes the prediction model error to accumulate, which also contributes to higher computation times as the reference trajectory may become difficult for the prediction model to track. Therefore, the suitable prediction horizon for tracking, in terms of acceptable accuracy for lower computation times, is 10 or 15.

%During this maneuver, similar to NL--5, the MILP controllers all show the same accuracy for different $\Np$ values. However, NL--1 exhibits oscillatory behavior in terms of tracking accuracy across the $\Np$ axis. This is due to the fact that increasing the prediction horizon, even by using the same prediction model as the ground truth, increases the dimensions of the search space, hence decreasing the probability of converging to an ``acceptable" optimum when the problem is only solved via the warm-start strategy. Performing the aggressive lane change maneuver as shown in Fig.~\ref{fig:mans} requires drastic changes in the input signals (compared to its safer counterpart) where relying on the warm-start strategy limits the chance of converging to a optimum with an acceptable tracking error.

In terms of computation speed, the MILP controllers with the T model show a steady increase rate similar to the nonlinear ideal models while other hybrid controllers show an increase in the rate of computation time after a certain $\Np$ value, which stems from the prediction model accuracy. Longer prediction horizons do not only increase the dimensions of the search space in the optimization problem, but also lead to accumulation of the prediction model error. This accumulation increases the error as well as computation time as the reference trajectory can become infeasible to track for the prediction model, leading to slower convergence. Therefore, the T model yields the best accuracy compared to the R and S models.

Controllers with the T model in general show the lowest increase of computation time when compared to the other hybrid controllers. In the MIQCP controllers T--REL and T--BEL the rate of increase is higher for $\Np$ between 5 and 10, but it converges to the same rate as the NLP controllers and their MILP counterparts T--RMP and T--BMP. This stems from the limitations of using the ellipsoidal constraints compared to the polytopic ones: the ellipsoidal approximations of the feasible region have a lower coverage of the feasible region (for more details, see Table~V in Part~I of this paper), which limits using the full control potential for an aggressive maneuver in shorter prediction horizons.

As Fig.~\ref{fig:ideal2} shows, the MILP controllers exceed the performance of the MIQCP ones in terms of accuracy, as well as computation speed. Therefore, for the next simulations we only consider the MILP controllers as prospective suitable hybrid candidates.

\subsubsection{Cornering Maneuvers}\input{Plot_IdealMan345}

We compare the MILP and NLP controllers during the three cornering maneuvers shown in Fig.~\ref{fig:ideal345}. In maneuvers 3, and 5 where the input forces vary drastically over the maneuver (see Fig.~\ref{fig:mans}), the NL--1 controller shows a poor computational performance and oscillatory behavior in the error plots across the $\Np$ axis, which was also observed in the aggressive lane change maneuver and discussed there.

In all three cornering maneuvers, the controller with the T model yields the best computational performance with its mean tracking error below 4.5\% and maximum error below 10\% in all cases.

Just as for the lane change maneuvers, increasing $\Np$ leads to a higher computation time for all the controllers; however, the rate of increase is the lowest for the nonlinear MPC and the T--BMP controller. For the T--RMP controller the same behavior is observed for \mbox{$\Np<20$}. Comparing the performance of the hybrid and nonlinear MPC controllers in all the five maneuvers, a suitable prediction horizon for tracking, in terms of acceptable accuracy for acceptable computation times, is 10 or 15. Next, we select $\Np = 10$ for the comparison of the robustness of the controllers to friction uncertainty. However, we also have simulated other $\Np$ values, reaching similar results. Therefore, for a compact presentation, we present the trends and analyze them for $\Np = 10$. In addition, since the MILP controllers with the S prediction model show larger tracking errors and larger computation times, especially for shorter prediction horizons, we disregard them at this stage and compare the four MILP controllers with R and T prediction models against their NLP counterparts.

\begin{rem}
	We have simulated the scenarios for other $N_{\rm{p}}$ values and reached similar results. Therefore, for a compact presentation, we present the trends and analyze them for \mbox{$\Np = 10$} as a representative value.
\end{rem}

\subsection{Friction Offset}\label{sec:submuoff}\begin{figure*}[hbtp]
\begin{center}
\begin{subfigure}[t]{\linewidth}
	\begin{tikzpicture}
	\begin{semilogyaxis}[width=0.3\textwidth,height=0.2\textwidth,	
		xlabel={$\mu$},ylabel={Mean Rel. Err. (\%)},
		xmin=0.68, xmax=1.02,ymin=0, ymax=26,
		xtick={0.5,0.6,0.7,0.8,0.9,1.0},ytick={2,5,10},scaled ticks=false,tick label style={/pgf/number format/fixed},log ticks with fixed point,xmajorgrids=true, xminorgrids=true,ymajorgrids=true,yminorgrids=true,grid style=dashed]	
		\addplot[smooth,color=NL1,thick,mark=*] table [x=MU,y=NL1_MEAN]{Files/OFFSET_MAN2.dat};	
		\addplot[smooth,color=NL5,thick,mark=*] table [x=MU,y=NL5_MEAN]{Files/OFFSET_MAN2.dat};	
		\addplot[smooth,color=RRMP,thick,mark=*] table [x=MU,y=RRMP_MEAN]{Files/OFFSET_MAN2.dat};	
		\addplot[smooth,color=RBMP,thick,mark=*] table [x=MU,y=RBMP_MEAN]{Files/OFFSET_MAN2.dat};
		\addplot[smooth,color=TRMP,thick,mark=*] table [x=MU,y=TRMP_MEAN]{Files/OFFSET_MAN2.dat};	
		\addplot[smooth,color=TBMP,thick,mark=*] table [x=MU,y=TBMP_MEAN]{Files/OFFSET_MAN2.dat};	
	\end{semilogyaxis}
	\end{tikzpicture}
	\begin{tikzpicture}
	\begin{semilogyaxis}[width=0.3\textwidth,height=0.2\textwidth,	
		xlabel={$\mu$},ylabel={Max Rel. Err. (\%)},
		xmin=0.68, xmax=1.02,ymin=0, ymax=78,
		xtick={0.5,0.6,0.7,0.8,0.9,1.0},ytick={2,5,10,25,50},scaled ticks=false,tick label style={/pgf/number format/fixed},log ticks with fixed point,xmajorgrids=true, xminorgrids=true,ymajorgrids=true,yminorgrids=true,grid style=dashed]	
		\addplot[smooth,color=NL1,thick,mark=*] table [x=MU,y=NL1_MAX]{Files/OFFSET_MAN2.dat};	
		\addplot[smooth,color=NL5,thick,mark=*] table [x=MU,y=NL5_MAX]{Files/OFFSET_MAN2.dat};	
		\addplot[smooth,color=RRMP,thick,mark=*] table [x=MU,y=RRMP_MAX]{Files/OFFSET_MAN2.dat};	
		\addplot[smooth,color=RBMP,thick,mark=*] table [x=MU,y=RBMP_MAX]{Files/OFFSET_MAN2.dat};
		\addplot[smooth,color=TRMP,thick,mark=*] table [x=MU,y=TRMP_MAX]{Files/OFFSET_MAN2.dat};	
		\addplot[smooth,color=TBMP,thick,mark=*] table [x=MU,y=TBMP_MAX]{Files/OFFSET_MAN2.dat};	
	\end{semilogyaxis}
	\end{tikzpicture}
	\begin{tikzpicture}
	\begin{semilogyaxis}[width=0.3\textwidth,height=0.2\textwidth,	
		xlabel={$\mu$},ylabel={Comp. Time (s)},
		xmin=0.68, xmax=1.02,ymin=0.05, ymax=30,
		xtick={0.5,0.6,0.7,0.8,0.9,1.0},ytick={0.1,1.0,10},xmajorgrids=true, xminorgrids=true,ymajorgrids=true,yminorgrids=true,grid style=dashed,legend pos=outer north east]
		\addplot[smooth,color=NL1,thick,mark=*] table [x=MU,y=NL1_TIME]{Files/OFFSET_MAN2.dat};	
		\addplot[smooth,color=NL5,thick,mark=*] table [x=MU,y=NL5_TIME]{Files/OFFSET_MAN2.dat};	
		\addplot[smooth,color=RRMP,thick,mark=*] table [x=MU,y=RRMP_TIME]{Files/OFFSET_MAN2.dat};	
		\addplot[smooth,color=RBMP,thick,mark=*] table [x=MU,y=RBMP_TIME]{Files/OFFSET_MAN2.dat};
		\addplot[smooth,color=TRMP,thick,mark=*] table [x=MU,y=TRMP_TIME]{Files/OFFSET_MAN2.dat};	
		\addplot[smooth,color=TBMP,thick,mark=*] table [x=MU,y=TBMP_TIME]{Files/OFFSET_MAN2.dat};		
		\legend{NL--1, NL--5, R--RMP, R--BMP, T--RMP, T--BMP}
	\end{semilogyaxis}
	\end{tikzpicture}
	\subcaption{Aggressive lane change maneuver (maneuver 2 in Table~\ref{tab:man})}
	\label{fig:offset2}
\end{subfigure}		
\begin{subfigure}[t]{\linewidth}
	\begin{tikzpicture}
	\begin{semilogyaxis}[width=0.3\textwidth,height=0.2\textwidth,	
		xlabel={$\mu$},ylabel={Mean Rel. Err. (\%)},
		xmin=0.68, xmax=1.02,ymin=0, ymax=26,
		xtick={0.5,0.6,0.7,0.8,0.9,1.0},ytick={2,5,10},scaled ticks=false,tick label style={/pgf/number format/fixed},log ticks with fixed point,xmajorgrids=true, xminorgrids=true,ymajorgrids=true,yminorgrids=true,grid style=dashed]	
		\addplot[smooth,color=NL1,thick,mark=*] table [x=MU,y=NL1_MEAN]{Files/OFFSET_MAN3.dat};	
		\addplot[smooth,color=NL5,thick,mark=*] table [x=MU,y=NL5_MEAN]{Files/OFFSET_MAN3.dat};	
		\addplot[smooth,color=RRMP,thick,mark=*] table [x=MU,y=RRMP_MEAN]{Files/OFFSET_MAN3.dat};	
		\addplot[smooth,color=RBMP,thick,mark=*] table [x=MU,y=RBMP_MEAN]{Files/OFFSET_MAN3.dat};
		\addplot[smooth,color=TRMP,thick,mark=*] table [x=MU,y=TRMP_MEAN]{Files/OFFSET_MAN3.dat};	
		\addplot[smooth,color=TBMP,thick,mark=*] table [x=MU,y=TBMP_MEAN]{Files/OFFSET_MAN3.dat};	
	\end{semilogyaxis}
	\end{tikzpicture}
	\begin{tikzpicture}
	\begin{semilogyaxis}[width=0.3\textwidth,height=0.2\textwidth,	
		xlabel={$\mu$},ylabel={Max Rel. Err. (\%)},
		xmin=0.68, xmax=1.02,ymin=0, ymax=78,
		xtick={0.5,0.6,0.7,0.8,0.9,1.0},ytick={2,5,10,25,50},scaled ticks=false,tick label style={/pgf/number format/fixed},log ticks with fixed point,xmajorgrids=true, xminorgrids=true,ymajorgrids=true,yminorgrids=true,grid style=dashed]	
		\addplot[smooth,color=NL1,thick,mark=*] table [x=MU,y=NL1_MAX]{Files/OFFSET_MAN3.dat};	
		\addplot[smooth,color=NL5,thick,mark=*] table [x=MU,y=NL5_MAX]{Files/OFFSET_MAN3.dat};	
		\addplot[smooth,color=RRMP,thick,mark=*] table [x=MU,y=RRMP_MAX]{Files/OFFSET_MAN3.dat};	
		\addplot[smooth,color=RBMP,thick,mark=*] table [x=MU,y=RBMP_MAX]{Files/OFFSET_MAN3.dat};
		\addplot[smooth,color=TRMP,thick,mark=*] table [x=MU,y=TRMP_MAX]{Files/OFFSET_MAN3.dat};	
		\addplot[smooth,color=TBMP,thick,mark=*] table [x=MU,y=TBMP_MAX]{Files/OFFSET_MAN3.dat};	
	\end{semilogyaxis}
	\end{tikzpicture}
	\begin{tikzpicture}
	\begin{semilogyaxis}[width=0.3\textwidth,height=0.2\textwidth,	
		xlabel={$\mu$},ylabel={Comp. Time (s)},
		xmin=0.68, xmax=1.02,ymin=0.05, ymax=30,
		xtick={0.5,0.6,0.7,0.8,0.9,1.0},ytick={0.1,1.0,10},
		scaled ticks=false,tick label style={/pgf/number format/fixed},xmajorgrids=true, xminorgrids=true,ymajorgrids=true,yminorgrids=true,grid style=dashed,legend pos=outer north east]
		\addplot[smooth,color=NL1,thick,mark=*] table [x=MU,y=NL1_TIME]{Files/OFFSET_MAN3.dat};	
		\addplot[smooth,color=NL5,thick,mark=*] table [x=MU,y=NL5_TIME]{Files/OFFSET_MAN3.dat};	
		\addplot[smooth,color=RRMP,thick,mark=*] table [x=MU,y=RRMP_TIME]{Files/OFFSET_MAN3.dat};	
		\addplot[smooth,color=RBMP,thick,mark=*] table [x=MU,y=RBMP_TIME]{Files/OFFSET_MAN3.dat};
		\addplot[smooth,color=TRMP,thick,mark=*] table [x=MU,y=TRMP_TIME]{Files/OFFSET_MAN3.dat};	
		\addplot[smooth,color=TBMP,thick,mark=*] table [x=MU,y=TBMP_TIME]{Files/OFFSET_MAN3.dat};		
		\legend{NL--1, NL--5, R--RMP, R--BMP, T--RMP, T--BMP}
	\end{semilogyaxis}
	\end{tikzpicture}
	\subcaption{Drift cornering maneuver (maneuver 3 in Table~\ref{tab:man})}
	\label{fig:offset3}
\end{subfigure}
\begin{subfigure}[t]{\linewidth}
	\begin{tikzpicture}
	\begin{semilogyaxis}[width=0.3\textwidth,height=0.2\textwidth,	
		xlabel={$\mu$},ylabel={Mean Rel. Err. (\%)},
		xmin=0.68, xmax=1.02,ymin=0, ymax=26,
		xtick={0.5,0.6,0.7,0.8,0.9,1.0},ytick={2,5,10},
		scaled ticks=false,tick label style={/pgf/number format/fixed},log ticks with fixed point,xmajorgrids=true, xminorgrids=true,ymajorgrids=true,yminorgrids=true,grid style=dashed]	
		\addplot[smooth,color=NL1,thick,mark=*] table [x=MU,y=NL1_MEAN]{Files/OFFSET_MAN5.dat};	
		\addplot[smooth,color=NL5,thick,mark=*] table [x=MU,y=NL5_MEAN]{Files/OFFSET_MAN5.dat};	
		\addplot[smooth,color=RRMP,thick,mark=*] table [x=MU,y=RRMP_MEAN]{Files/OFFSET_MAN5.dat};	
		\addplot[smooth,color=RBMP,thick,mark=*] table [x=MU,y=RBMP_MEAN]{Files/OFFSET_MAN5.dat};
		\addplot[smooth,color=TRMP,thick,mark=*] table [x=MU,y=TRMP_MEAN]{Files/OFFSET_MAN5.dat};	
		\addplot[smooth,color=TBMP,thick,mark=*] table [x=MU,y=TBMP_MEAN]{Files/OFFSET_MAN5.dat};	
	\end{semilogyaxis}
	\end{tikzpicture}
	\begin{tikzpicture}
	\begin{semilogyaxis}[width=0.3\textwidth,height=0.2\textwidth,	
		xlabel={$\mu$},ylabel={Max Rel. Err. (\%)},
		xmin=0.68, xmax=1.02,ymin=0, ymax=78,
		xtick={0.5,0.6,0.7,0.8,0.9,1.0},ytick={2,5,10,25,50},scaled ticks=false,tick label style={/pgf/number format/fixed},log ticks with fixed point,xmajorgrids=true, xminorgrids=true,ymajorgrids=true,yminorgrids=true,grid style=dashed]	
		\addplot[smooth,color=NL1,thick,mark=*] table [x=MU,y=NL1_MAX]{Files/OFFSET_MAN5.dat};	
		\addplot[smooth,color=NL5,thick,mark=*] table [x=MU,y=NL5_MAX]{Files/OFFSET_MAN5.dat};	
		\addplot[smooth,color=RRMP,thick,mark=*] table [x=MU,y=RRMP_MAX]{Files/OFFSET_MAN5.dat};	
		\addplot[smooth,color=RBMP,thick,mark=*] table [x=MU,y=RBMP_MAX]{Files/OFFSET_MAN5.dat};
		\addplot[smooth,color=TRMP,thick,mark=*] table [x=MU,y=TRMP_MAX]{Files/OFFSET_MAN5.dat};	
		\addplot[smooth,color=TBMP,thick,mark=*] table [x=MU,y=TBMP_MAX]{Files/OFFSET_MAN5.dat};	
	\end{semilogyaxis}
	\end{tikzpicture}
	\begin{tikzpicture}
	\begin{semilogyaxis}[width=0.3\textwidth,height=0.2\textwidth,	
		xlabel={$\mu$},ylabel={Comp. Time (s)},
		xmin=0.68, xmax=1.02,ymin=0.05, ymax=30,
		xtick={0.5,0.6,0.7,0.8,0.9,1.0},ytick={0.1,1.0,10},xmajorgrids=true, xminorgrids=true,ymajorgrids=true,yminorgrids=true,grid style=dashed,legend pos=outer north east]
		\addplot[smooth,color=NL1,thick,mark=*] table [x=MU,y=NL1_TIME]{Files/OFFSET_MAN5.dat};	
		\addplot[smooth,color=NL5,thick,mark=*] table [x=MU,y=NL5_TIME]{Files/OFFSET_MAN5.dat};	
		\addplot[smooth,color=RRMP,thick,mark=*] table [x=MU,y=RRMP_TIME]{Files/OFFSET_MAN5.dat};	
		\addplot[smooth,color=RBMP,thick,mark=*] table [x=MU,y=RBMP_TIME]{Files/OFFSET_MAN5.dat};
		\addplot[smooth,color=TRMP,thick,mark=*] table [x=MU,y=TRMP_TIME]{Files/OFFSET_MAN5.dat};	
		\addplot[smooth,color=TBMP,thick,mark=*] table [x=MU,y=TBMP_TIME]{Files/OFFSET_MAN5.dat};		
		\legend{NL--1, NL--5, R--RMP, R--BMP, T--RMP, T--BMP}
	\end{semilogyaxis}
	\end{tikzpicture}
	\subcaption{Low-speed cornering maneuver (maneuver 5 in Table~\ref{tab:man})}
	\label{fig:offset5}
\end{subfigure}		
\caption{Computational performance of the nonlinear and hybrid MPC controllers during three reference maneuvers (maneuvers 3, 4, and 5 in Table~\ref{tab:man}) in the friction offset scenario.}\label{fig:offset}
\end{center}
\end{figure*}
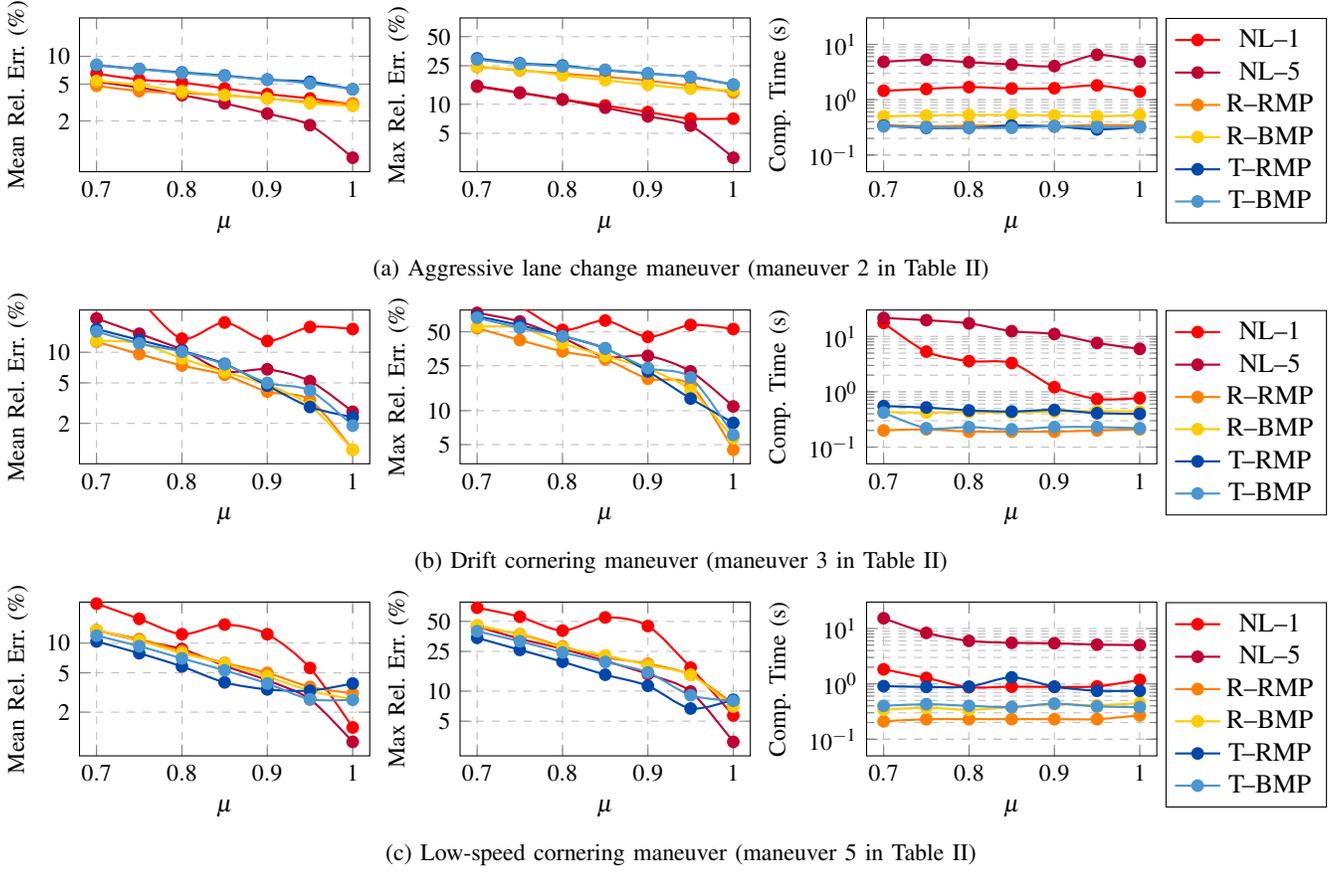

The maneuvers 2, 3, and 5 in Table~\ref{tab:man} are the three most critical ones: here, the vehicle operates close to the tire saturation and stability limits almost the whole time as shown in Fig.~\ref{fig:mans}. Thus, we used these maneuvers to study the effect of friction offset. The prediction horizon is selected as $\Np = 10$ and the simulations are run for different road friction coefficients in the range $\mu \in \{0.70, 0.75, \dots, 1.00\}$ to account for uncertain friction in the prediction model. Figure~\ref{fig:offset} shows the computational performance of the nonlinear and hybrid MPC controllers during the three reference maneuvers.

While the computation time for the hybrid controllers does not vary by increasing the friction uncertainty, the nonlinear controllers show an increase in the computation time in maneuvers 3 and 5 where a significant fraction of the maneuver is performed close to the tire-saturation and vehicle-handling limits, which are functions of the friction coefficient. 

The maximum and mean tracking errors increase for lower friction coefficients for all the controllers. However, the rate of error increase for the nonlinear controllers is higher. The difference between the tracking errors of NL--1 and NL--5 once again indicates the shortcoming of a warm-start strategy during aggressive maneuvers in searching for the optimal solution in the search space. This however comes at the price of an increase in computation times, best shown in Fig.~\ref{fig:offset5} where solving the NLP for five initial guesses increases the computation time tenfold.
\begin{rem}
	The reason behind the computational increase in NLP is as follows: compared to the shifted solution to the previous step, the other initial guesses are generally further away from a local optimum. As a result, the increase in computation time is more than linear.
\end{rem}
In the presence of friction offset, the tracking error of NL--5 converges to that of the MILP controllers in maneuvers 2, 3, and 5 where a more extensive search over the search space is required to perform the maneuver from an initial state with an error from the previous solution. To understand this phenomenon and its two contributing factors, we look at the NL--5 and T--BMP controllers in more detail. 

Notice that NL--5 and T--BMP have mean tracking errors below 5\% in all the maneuvers of Fig.~\ref{fig:offset} with $\mu = 1$, which generates the same friction as their prediction models. When $\mu$ on the road is decreased to $0.7$, the controllers still seek to find the solution (including tire forces) close to the boundary of the feasible region of their model, which assumes $\mu \approx 1$. However, these forces cannot be generated by the tire in the real system due to the lower friction on the road. Therefore, the first contributing factor to the error is the fact that error accumulation grows exponentially with the number of control time steps and as a result the controller converges to an infeasible solution for the real system (note that the real feasible region is shrinking with the friction reduction). Secondly, with larger errors, finding a feasible solution to track the reference trajectory from an initial state with an already large tracking error might not be possible after a certain error bound. This not only increases the convergence time for the NL--5 controller, but also results in converging to even worse solutions both in terms of constraint violation and optimality to the point where we observe the tracking error of NL--5 exceeds the error for T--BMP in Figures~\ref{fig:offset3} and~\ref{fig:offset5} as with a similar order of model error, the branch-and-bound approach of the MILP solver, as opposed to the NLP solver, guarantees convergence to the global optimum if given enough time, while keeping the computation time as low as for the ideal case.

\subsection{Friction Disturbance}\label{sec:submudist} \begin{figure*}[t]
\begin{subfigure}[t]{0.48\textwidth}\centering
	\begin{tikzpicture}
	\begin{semilogyaxis}[width=0.85\textwidth,height=0.45\textwidth,	
		xlabel={Time (s)},ylabel={Rel. Err. (\%)},
		xmin=0.00, xmax=2.0,ymin=0.5, ymax=120,
		xtick={0.0,0.50,1.0,1.5,2},ytick={1,5,25,100},scaled ticks=false,tick label style={/pgf/number format/fixed},log ticks with fixed point,xmajorgrids=true, xminorgrids=true,ymajorgrids=true,yminorgrids=true,grid style=dashed,legend pos=outer north east]
		\addplot[smooth,color=NL1,ultra thick] table [x=T,y=NL11]{Files/DISTURBANCE.dat};	
		\addplot[smooth,color=NL5,ultra thick] table [x=T,y=NL51]{Files/DISTURBANCE.dat};	
		\addplot[smooth,color=RRMP,ultra thick] table [x=T,y=RRMP1]{Files/DISTURBANCE.dat};	
		\addplot[smooth,color=RBMP,ultra thick] table [x=T,y=RBMP1]{Files/DISTURBANCE.dat};	
		\addplot[smooth,color=TRMP,ultra thick] table [x=T,y=TRMP1]{Files/DISTURBANCE.dat};	
		\addplot[smooth,color=TBMP,ultra thick] table [x=T,y=TBMP1]{Files/DISTURBANCE.dat};	
		\path[fill=orange,fill opacity=0.15] 
		(0.5, 0.01)--(0.5,120)--(1,120)--(1,0.01)--cycle;
		\node[text width=1.5cm] at (0.82,60) {$\mu = 0.4$};
		%\legend{1.68s, 6.39s, 0.31s, 0.43s, 0.31s, 0.26s}
		\legend{NL--1,NL--5,R--RMP,R--BMP,T--RMP,T--BMP}
	\end{semilogyaxis}
	\end{tikzpicture}
	\subcaption{Safe lane change maneuver (maneuver 1 in Table~\ref{tab:man})}
	\label{fig:dist1}
	\end{subfigure}
\begin{subfigure}[t]{0.48\textwidth}\centering
	\begin{tikzpicture}
	\begin{semilogyaxis}[width=0.85\textwidth,height=0.45\textwidth,	
		xlabel={Time (s)},ylabel={Rel. Err. (\%)},
		xmin=0.00, xmax=2.0,ymin=0.5, ymax=120,
		xtick={0.0,0.50,1.0,1.5,2},ytick={1,5,25,100},scaled ticks=false,tick label style={/pgf/number format/fixed},log ticks with fixed point,xmajorgrids=true, xminorgrids=true,ymajorgrids=true,yminorgrids=true,grid style=dashed,legend pos=outer north east]
		\addplot[smooth,color=NL1,ultra thick] table [x=T,y=NL12]{Files/DISTURBANCE.dat};	
		\addplot[smooth,color=NL5,ultra thick] table [x=T,y=NL52]{Files/DISTURBANCE.dat};	
		\addplot[smooth,color=RRMP,ultra thick] table [x=T,y=RRMP2]{Files/DISTURBANCE.dat};	
		\addplot[smooth,color=RBMP,ultra thick] table [x=T,y=RBMP2]{Files/DISTURBANCE.dat};	
		\addplot[smooth,color=TRMP,ultra thick] table [x=T,y=TRMP2]{Files/DISTURBANCE.dat};	
		\addplot[smooth,color=TBMP,ultra thick] table [x=T,y=TBMP2]{Files/DISTURBANCE.dat};	
		\path[fill=orange,fill opacity=0.15] 
		(0.5, 0.01)--(0.5,120)--(1,120)--(1,0.01)--cycle;
		\node[text width=1.5cm] at (0.82,35) {$\mu = 0.4$};
		%\legend{1.49s, 4.68s, 0.31s, 0.45s, 0.33s, 0.34s}
		\legend{NL--1,NL--5,R--RMP,R--BMP,T--RMP,T--BMP}
	\end{semilogyaxis}
	\end{tikzpicture}
	\subcaption{Aggressive lane change maneuver (maneuver 2 in Table~\ref{tab:man})}
	\label{fig:dist2}
\end{subfigure}		
\begin{subfigure}[t]{0.48\textwidth}\centering
	\begin{tikzpicture}
	\begin{semilogyaxis}[width=0.85\textwidth,height=0.45\textwidth,	
		xlabel={Time (s)},ylabel={Rel. Err. (\%)},
		xmin=0.00, xmax=2.0,ymin=0.5, ymax=120,
		xtick={0.0,0.50,1.0,1.5,2},ytick={1,5,25,100},scaled ticks=false,tick label style={/pgf/number format/fixed},log ticks with fixed point,xmajorgrids=true, xminorgrids=true,ymajorgrids=true,yminorgrids=true,grid style=dashed,legend pos=outer north east]
		\addplot[smooth,color=NL1,ultra thick] table [x=T,y=NL13]{Files/DISTURBANCE.dat};	
		\addplot[smooth,color=NL5,ultra thick] table [x=T,y=NL53]{Files/DISTURBANCE.dat};	
		\addplot[smooth,color=RRMP,ultra thick] table [x=T,y=RRMP3]{Files/DISTURBANCE.dat};	
		\addplot[smooth,color=RBMP,ultra thick] table [x=T,y=RBMP3]{Files/DISTURBANCE.dat};	
		\addplot[smooth,color=TRMP,ultra thick] table [x=T,y=TRMP3]{Files/DISTURBANCE.dat};	
		\addplot[smooth,color=TBMP,ultra thick] table [x=T,y=TBMP3]{Files/DISTURBANCE.dat};	
		\path[fill=orange,fill opacity=0.15] 
		(0.5, 0.01)--(0.5,120)--(1,120)--(1,0.01)--cycle;
		\node[text width=1.5cm] at (0.82,1) {$\mu = 0.4$};
		%\legend{0.39s, 4.50s, 0.20s, 0.43s, 0.43s, 0.24s}
		\legend{NL--1,NL--5,R--RMP,R--BMP,T--RMP,T--BMP}
	\end{semilogyaxis}
	\end{tikzpicture}
	\subcaption{Drift cornering maneuver (maneuver 3 in Table~\ref{tab:man})}
	\label{fig:dist3}
\end{subfigure}
\begin{subfigure}[t]{0.48\textwidth}\centering
	\begin{tikzpicture}
	\begin{semilogyaxis}[width=0.85\textwidth,height=0.45\textwidth,	
		xlabel={Time (s)},ylabel={Rel. Err. (\%)},
		xmin=0.00, xmax=2.0,ymin=0.5, ymax=120,
		xtick={0.0,0.50,1.0,1.5,2},ytick={1,5,25,100},scaled ticks=false,tick label style={/pgf/number format/fixed},log ticks with fixed point,xmajorgrids=true, xminorgrids=true,ymajorgrids=true,yminorgrids=true,grid style=dashed,legend pos=outer north east]
		\addplot[smooth,color=NL1,ultra thick] table [x=T,y=NL14]{Files/DISTURBANCE.dat};	
		\addplot[smooth,color=NL5,ultra thick] table [x=T,y=NL54]{Files/DISTURBANCE.dat};	
		\addplot[smooth,color=RRMP,ultra thick] table [x=T,y=RRMP4]{Files/DISTURBANCE.dat};	
		\addplot[smooth,color=RBMP,ultra thick] table [x=T,y=RBMP4]{Files/DISTURBANCE.dat};	
		\addplot[smooth,color=TRMP,ultra thick] table [x=T,y=TRMP4]{Files/DISTURBANCE.dat};	
		\addplot[smooth,color=TBMP,ultra thick] table [x=T,y=TBMP4]{Files/DISTURBANCE.dat};	
		\path[fill=orange,fill opacity=0.15] 
		(0.5, 0.01)--(0.5,120)--(1,120)--(1,0.01)--cycle;
		\node[text width=1.5cm] at (0.82,35) {$\mu = 0.4$};
		%\legend{1.26s, 5.85s, 0.23s, 0.16s, 0.43s, 0.19s}
		\legend{NL--1,NL--5,R--RMP,R--BMP,T--RMP,T--BMP}
	\end{semilogyaxis}
	\end{tikzpicture}
	\subcaption{High-speed cornering maneuver (maneuver 4 in Table~\ref{tab:man})}
	\label{fig:dist4}
\end{subfigure}
\begin{subfigure}[t]{0.48\textwidth}\centering
	\begin{tikzpicture}
	\begin{semilogyaxis}[width=0.85\textwidth,height=0.45\textwidth,	
		xlabel={Time (s)},ylabel={Rel. Err. (\%)},
		xmin=0.00, xmax=2.0,ymin=0.5, ymax=120,
		xtick={0.0,0.50,1.0,1.5,2},ytick={1,5,25,100},scaled ticks=false,tick label style={/pgf/number format/fixed},log ticks with fixed point,xmajorgrids=true, xminorgrids=true,ymajorgrids=true,yminorgrids=true,grid style=dashed,legend pos=outer north east]
		\addplot[smooth,color=NL1,ultra thick] table [x=T,y=NL15]{Files/DISTURBANCE.dat};	
		\addplot[smooth,color=NL5,ultra thick] table [x=T,y=NL55]{Files/DISTURBANCE.dat};	
		\addplot[smooth,color=RRMP,ultra thick] table [x=T,y=RRMP5]{Files/DISTURBANCE.dat};	
		\addplot[smooth,color=RBMP,ultra thick] table [x=T,y=RBMP5]{Files/DISTURBANCE.dat};	
		\addplot[smooth,color=TRMP,ultra thick] table [x=T,y=TRMP5]{Files/DISTURBANCE.dat};	
		\addplot[smooth,color=TBMP,ultra thick] table [x=T,y=TBMP5]{Files/DISTURBANCE.dat};	
		\path[fill=orange,fill opacity=0.15] 
		(0.5, 0.01)--(0.5,120)--(1,120)--(1,0.01)--cycle;
		\node[text width=1.5cm] at (0.82,1) {$\mu = 0.4$};
		%\legend{7.33s, 20.32s, 0.24s, 0.37s, 0.97s, 0.93s}
		\legend{NL--1,NL--5,R--RMP,R--BMP,T--RMP,T--BMP}
	\end{semilogyaxis}
	\end{tikzpicture}
	\subcaption{Low-speed cornering maneuver (maneuver 5 in Table~\ref{tab:man})}
	\label{fig:dist5}
\end{subfigure}		
\begin{subfigure}[t]{0.57\textwidth}\centering
	\begin{tikzpicture}
	\begin{axis}[ybar,width=0.7\textwidth,height=0.375\textwidth,legend pos=outer north east,
		ymin=0.1, ymax=20,bar width=2pt,symbolic x coords={M1,M2,M3,M4,M5},xtick align=inside,
		xlabel=Reference Maneuver,ylabel=Mean Comp. Time (s),ymode = log, log origin=infty, xminorgrids=true,ymajorgrids=true,grid style=dashed,xtick=data]
		\addplot[fill=NL1,color=NL1] coordinates {(M1,1.68)(M2,1.49)(M3,0.39)(M4,1.26)(M5,7.33)};
		\addplot[fill=NL5,color=NL5] coordinates {(M1,6.39)(M2,4.68)(M3,4.50)(M4,5.85)(M5,20.32)};
		\addplot[fill=RRMP,color=RRMP] coordinates {(M1,0.31)(M2,0.31)(M3,0.20)(M4,0.23)(M5,0.24)};		
		\addplot[fill=RBMP,color=RBMP] coordinates {(M1,0.43)(M2,0.45)(M3,0.43)(M4,0.16)(M5,0.37)};		
		\addplot[fill=TRMP,color=TRMP] coordinates {(M1,0.31)(M2,0.33)(M3,0.43)(M4,0.43)(M5,0.97)};		
		\addplot[fill=TBMP,color=TBMP] coordinates {(M1,0.26)(M2,0.34)(M3,0.24)(M4,0.19)(M5,0.93)};
		\legend{NL--1,NL--5,R--RMP,R--BMP,T--RMP,T--BMP}
	\end{axis}
	\end{tikzpicture}
	\subcaption{Computation times}
	\label{fig:disttimes}
\end{subfigure}		
\caption{Tracking error of the nonlinear and four selected MILP MPC controllers during five reference maneuvers in (\subref{fig:dist1}) to (\subref{fig:dist5}) in case of friction disturbance. The average computation time for each control time step is shown in (\subref{fig:disttimes}).}\label{fig:dist}
\end{figure*}
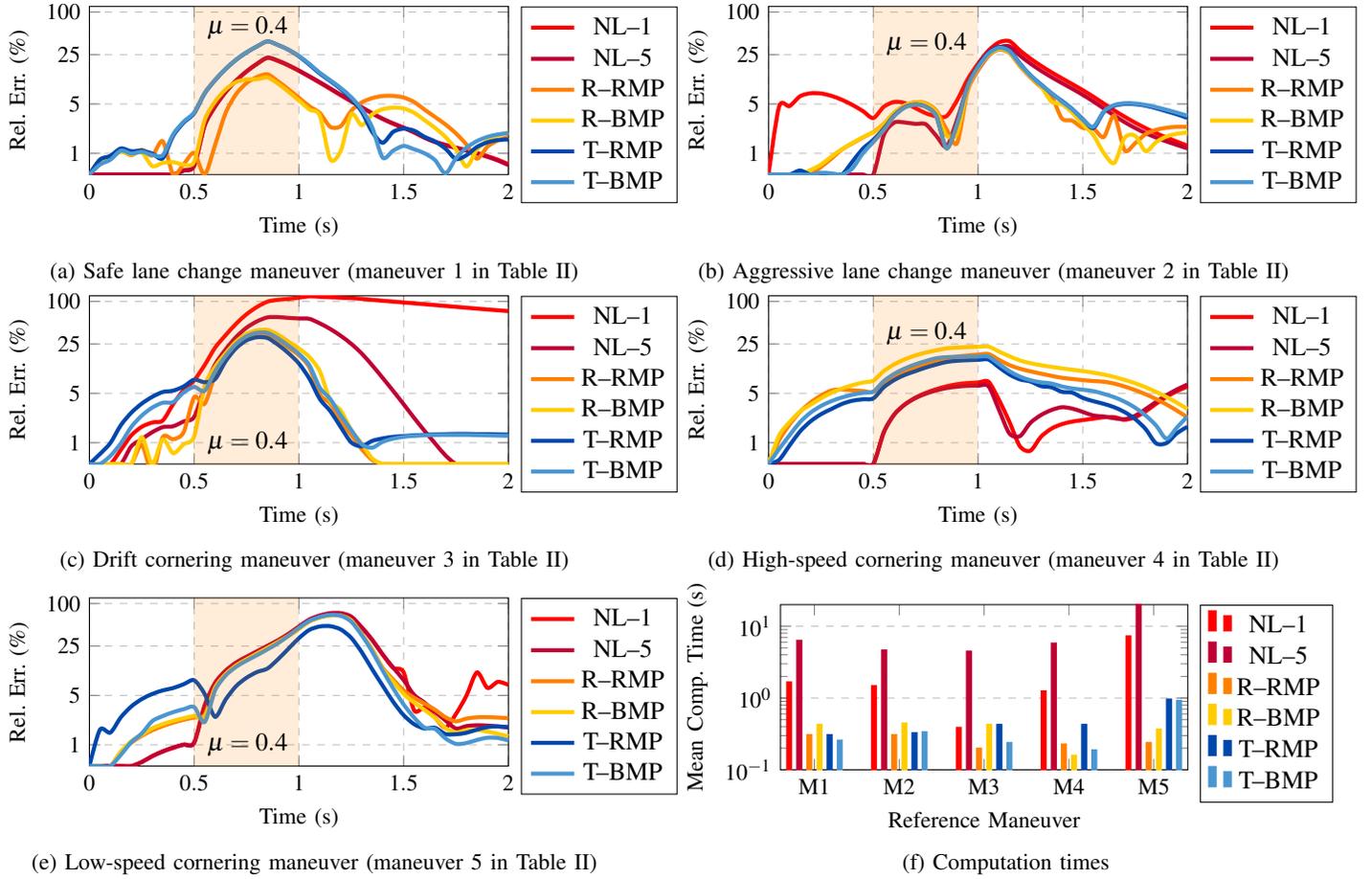

In this case, we assume a correct knowledge of the road friction during the maneuver, while exposing the system to a sudden friction reduction to $\mu = 0.4$ during the second quarter of the maneuver and we evaluate the tracking error, computation times, and the recovering ability of the closed-loop system without any estimation or corrections during the disturbance.

The tracking errors of the controllers are compared in case of sudden reduction of the friction to $\mu = 0.4$ in the second quarter of the maneuver to represent a case similar to pouring water on the road surface. The tracking errors at each time step are shown in Fig.~\ref{fig:dist} to compare the errors, as well as the ability of the different controllers in recovering from the friction disturbance. The average computation times per control step for each controller are presented in a separate plot in Fig.~\ref{fig:disttimes}.

During the safe lane change maneuver, all controllers recover to a tracking error below 5\% after five time steps, while the tracking error is larger during the other maneuvers that are more aggressive. After the friction disturbance in maneuvers 2 and 5, the tracking error keeps on increasing as the controllers fail to recover to an acceptable error bound. This could be understood by taking into account that the second quarter of maneuver 2 is when tracking the reference trajectory requires tire forces that will no longer be feasible due to the fact that the radius of the Kamm circle for the rear tire has decreased to 40\% of its original value, which means that following the planned trajectory will no longer be feasible for the prediction model. The same issue arises during maneuver 5 that can be tracked provided that the rear tire is generating forces close to its saturation limit during the whole maneuver (Fig.~\ref{fig:mankamm}).

The hybrid controllers show smaller tracking errors after recovery during the low-speed cornering maneuver while starting with larger tracking errors before the friction disturbance. This reduction is understandable in light of the fact that the second half of this maneuver requires the vehicle to operate further from the stability boundaries in $\beta$-$r$ envelope and the g-g diagram in Fig.~\ref{fig:mans}. Meanwhile, the tracking error of NL--1 stays above 5\%, which shows the limitations of depending on a warm-start strategy in convergence to better optima. This limitation is even more clear in Fig.~\ref{fig:dist3} for the drift cornering maneuver: while NL--5 recovers better after the disturbance, it still fails to reach smaller tracking errors as fast as the hybrid MPC controllers, which suggests that increasing the number of initial guesses by considering more than five points could improve the convergence capability of NL--5 to acceptable optima. However, it should be noted that even for five initial guesses, NL--5 requires 10 times more than the slowest hybrid MPC to converge to its best solution. 

In the safe lane change maneuver, T--RMP and T--BMP show larger tracking errors compared to the other controllers, and this is the scenario where the hybrid controllers in general show the highest error increase of 12\% to 30\% for 95\% computation time reduction. However, the effectiveness of the hybrid MPC controllers in terms of tracking error is more observable in more aggressive maneuvers where T--RMP and T--BMP show a better performance, in some cases even better than that of NL--1 and NL--5, due to the fact that the shortcomings of convergence to local optima is more clear in hazardous scenarios where there are sudden changes in the environment that require a more thorough search across the decision space. Comparing the control performance vs.\ computation time trade-off during the four aggressive reference maneuvers shows that choosing the hybrid MPC controllers R--BMP and R--RMP decreases the computation time to 2 to 5\%, while it increases the maximum error from 5\% to 15\% in maneuver 5 while decreasing it during maneuvers 2, 3, and 4 compared to NL--5. 

\subsection{Analysis of Computational Performance}

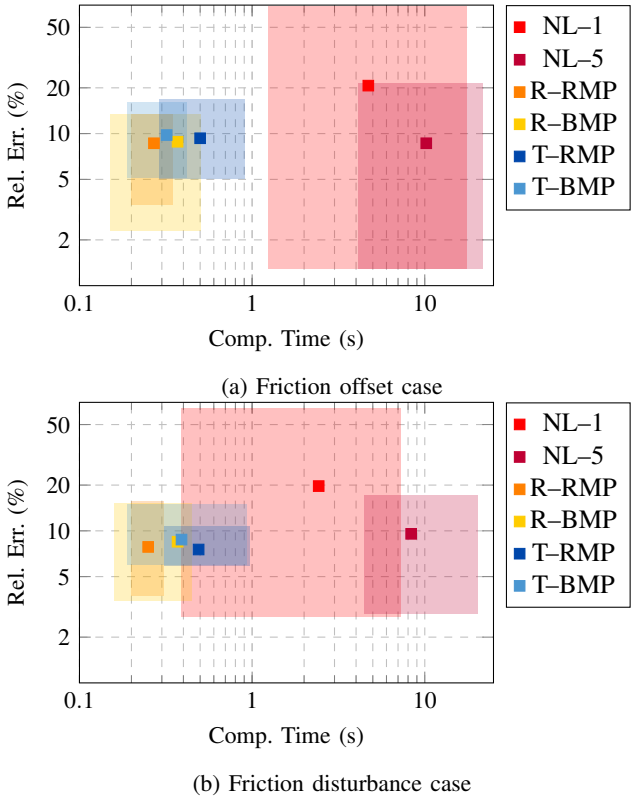
\begin{figure}[hbtp]
\begin{center}
\begin{subfigure}[t]{\linewidth}
	\begin{tikzpicture}
	\begin{loglogaxis}[width=0.8\textwidth,height=0.6\textwidth,	
		xlabel={Comp. Time (s)},ylabel={Rel. Err. (\%)},
		xmin=0.1,xmax=25,ymin=1,ymax=70,ytick={2,5,10,20,50},scaled ticks=false,tick label style={/pgf/number format/fixed},log ticks with fixed point,xmajorgrids=true, xminorgrids=true,ymajorgrids=true,yminorgrids=true,grid style=dashed,legend pos=outer north east]
		\path[fill=NL1,fill opacity=0.25] (1.24,1.3)--(1.24,68.9)--(17.55,68.9)--(17.55,1.3)--cycle;
		\addplot[only marks,mark=square*,NL1] coordinates {(4.71,20.68)};
		\path[fill=NL5,fill opacity=0.25] (4.1,1.3)--(4.1,21.5)--(21.58,21.5)--(21.58,1.3)--cycle;
		\addplot[only marks,mark=square*,NL5] coordinates {(10.20,8.64)};		
		\path[fill=RRMP,fill opacity=0.25] (0.2,3.4)--(0.2,13.3)--(0.35,13.3)--(0.35,3.4)--cycle;
		\addplot[only marks,mark=square*,RRMP] coordinates {(0.27,8.64)};
		\path[fill=RBMP,fill opacity=0.25] (0.15,2.3)--(0.15,13.5)--(0.50,13.5)--(0.50,2.3)--cycle;
		\addplot[only marks,mark=square*,RBMP] coordinates {(0.37,8.84)};		
		\path[fill=TRMP,fill opacity=0.25] (0.29,5.1)--(0.29,16.9)--(0.91,16.9)--(0.91,5.1)--cycle;
		\addplot[only marks,mark=square*,TRMP] coordinates {(0.50,9.32)};
		\path[fill=TBMP,fill opacity=0.25] (0.19,5.1)--(0.19,16.0)--(0.42,16.0)--(0.42,5.1)--cycle;
		\addplot[only marks,mark=square*,TBMP] coordinates {(0.32,9.78)};	
		\legend{NL--1,NL--5,R--RMP,R--BMP,T--RMP,T--BMP}	
	\end{loglogaxis}
	\end{tikzpicture}
	\subcaption{Friction offset case}
	\label{fig:compoffset}
\end{subfigure}			
\begin{subfigure}[t]{\linewidth}
	\begin{tikzpicture}
	\begin{loglogaxis}[width=0.8\textwidth,height=0.6\textwidth,	
		xlabel={Comp. Time (s)},ylabel={Rel. Err. (\%)},
		xmin=0.1,xmax=25,ymin=1,ymax=70,ytick={2,5,10,20,50},scaled ticks=false,tick label style={/pgf/number format/fixed},log ticks with fixed point,xmajorgrids=true, xminorgrids=true,ymajorgrids=true,yminorgrids=true,grid style=dashed,legend pos=outer north east]
		\path[fill=NL1,fill opacity=0.25] (0.39,2.74)--(0.39,64.21)--(7.33,64.21)--(7.33,2.74)--cycle;
		\addplot[only marks,mark=square*,NL1] coordinates {(2.43,19.7)};
		\path[fill=NL5,fill opacity=0.25] (4.5,2.83)--(4.5,17.13)--(20.32,17.13)--(20.32,2.83)--cycle;
		\addplot[only marks,mark=square*,NL5] coordinates {(8.34,9.56)};		
		\path[fill=RRMP,fill opacity=0.25] (0.2,3.74)--(0.2,15.61)--(0.31,15.61)--(0.31,3.74)--cycle;
		\addplot[only marks,mark=square*,RRMP] coordinates {(0.25,7.84)};
		\path[fill=RBMP,fill opacity=0.25] (0.16,3.49)--(0.16,15.33)--(0.45,15.33)--(0.45,3.49)--cycle;
		\addplot[only marks,mark=square*,RBMP] coordinates {(0.37,8.48)};		
		\path[fill=TRMP,fill opacity=0.25] (0.31,10.83)--(0.31,5.92)--(0.97,5.92)--(0.97,10.83)--cycle;
		\addplot[only marks,mark=square*,TRMP] coordinates {(0.49,7.54)};
		\path[fill=TBMP,fill opacity=0.25] (0.19,6.00)--(0.19,15.08)--(0.93,15.08)--(0.93,6.00)--cycle;
		\addplot[only marks,mark=square*,TBMP] coordinates {(0.39,8.76)};	
		\legend{NL--1,NL--5,R--RMP,R--BMP,T--RMP,T--BMP}	
	\end{loglogaxis}
	\end{tikzpicture}
	\subcaption{Friction disturbance case}
	\label{fig:compdist}
\end{subfigure}	
\caption{Relative errors and computation times for the nonlinear and four MILP MPC controllers during the five reference maneuvers in cases with friction uncertainty.}\label{fig:comp}
\end{center}
\end{figure}

Figure~\ref{fig:comp} plots the range of tracking error and computation times for the NLP and four MILP MPC controllers during all the five reference maneuvers in both friction offset and friction disturbance scenarios as shaded boxes. The average points are shown by square markings in the shaded area. Comparing the computational performance in Fig.~\ref{fig:comp} shows the power of the hybrid MPC controllers compared to the nonlinear ones in the presence of uncertainty. While the NLP controllers have a lower optimum, their maximum tracking errors reach much higher values while taking more time to converge. In terms of the average points, not only the best MILP controller in the ideal case (T--BMP) has a lower maximum error in both friction uncertainty cases compared to the best NLP one (NL--5), but it also has a higher computational efficiency: in the friction offset case, it trades off an error increase from 8.7\% to 9.7\% for reducing the computational time from 10.2s to 0.3s, and in the disturbance case it reaches a smaller tracking error (from 9.5\% to 8.7\%) as well as a lower computation time (from 8.3s to 0.4s).  

\subsection{Handling Limits}\label{sec:sublimit}\begin{figure}[hbtp]
\begin{center}
\begin{tikzpicture}
	\begin{semilogyaxis}[width=0.4\textwidth,height=0.25\textwidth,	
		xlabel={$\sqrt{a_x^2+a_y^2}$ (m/s$^2$)},ylabel={Mean Rel. Err. (\%)},
		xmin=4.8, xmax=8.6,ymin=0.5, ymax=35,
		xtick={5.33,6.19,7.14,8.21},ytick={2,5,10,25},
		scaled ticks=false,tick label style={/pgf/number format/fixed},log ticks with fixed point,xmajorgrids=true, xminorgrids=true,ymajorgrids=true,yminorgrids=true,grid style=dashed,legend columns=3,legend style={at={(0.5,1.5)},anchor=north}]	
		\addplot[smooth,color=NL1,thick,mark=*] table [x=GG,y=NL1_MEAN]{Files/AGGRESSIVE.dat};	
		\addplot[smooth,color=RRMP,thick,mark=*] table [x=GG,y=RRMP_MEAN]{Files/AGGRESSIVE.dat};	
		\addplot[smooth,color=TRMP,thick,mark=*] table [x=GG,y=TRMP_MEAN]{Files/AGGRESSIVE.dat};		
		\addplot[smooth,color=NL5,thick,mark=*] table [x=GG,y=NL5_MEAN]{Files/AGGRESSIVE.dat};			
		\addplot[smooth,color=RBMP,thick,mark=*] table [x=GG,y=RBMP_MEAN]{Files/AGGRESSIVE.dat};		
		\addplot[smooth,color=TBMP,thick,mark=*] table [x=GG,y=TBMP_MEAN]{Files/AGGRESSIVE.dat};	
		\addplot[semithick,dash pattern=on 2mm off 1mm,violet] coordinates{(8.3,0.1) (8.3,35)} node [pos=0.4,anchor=north, font=\footnotesize, sloped] {$\mu g$};
		\legend{NL--1, R--RMP, T--RMP, NL--5, R--BMP, T--BMP}
	\end{semilogyaxis}
\end{tikzpicture}
\begin{tikzpicture}
	\begin{semilogyaxis}[width=0.4\textwidth,height=0.25\textwidth,	
		xlabel={$\sqrt{a_x^2+a_y^2}$ (m/s$^2$)},ylabel={Max Rel. Err. (\%)},
		xmin=4.8, xmax=8.6,ymin=0.5, ymax=110,
		xtick={5.33,6.19,7.14,8.21},ytick={2,5,20,50},scaled ticks=false,tick label style={/pgf/number format/fixed},log ticks with fixed point,xmajorgrids=true, xminorgrids=true,ymajorgrids=true,yminorgrids=true,grid style=dashed]	
		\addplot[smooth,color=NL1,thick,mark=*] table [x=GG,y=NL1_MAX]{Files/AGGRESSIVE.dat};	
		\addplot[smooth,color=NL5,thick,mark=*] table [x=GG,y=NL5_MAX]{Files/AGGRESSIVE.dat};	
		\addplot[smooth,color=RRMP,thick,mark=*] table [x=GG,y=RRMP_MAX]{Files/AGGRESSIVE.dat};	
		\addplot[smooth,color=RBMP,thick,mark=*] table [x=GG,y=RBMP_MAX]{Files/AGGRESSIVE.dat};
		\addplot[smooth,color=TRMP,thick,mark=*] table [x=GG,y=TRMP_MAX]{Files/AGGRESSIVE.dat};	
		\addplot[smooth,color=TBMP,thick,mark=*] table [x=GG,y=TBMP_MAX]{Files/AGGRESSIVE.dat};	
		\addplot[semithick,dash pattern=on 2mm off 1mm,violet] coordinates{(8.3,0.1) (8.3,110)} node [pos=0.4,anchor=north, font=\footnotesize, sloped] {$\mu g$};
	\end{semilogyaxis}
\end{tikzpicture}
\begin{tikzpicture}
	\begin{semilogyaxis}[width=0.4\textwidth,height=0.25\textwidth,	
		xlabel={$\sqrt{a_x^2+a_y^2}$ (m/s$^2$)},ylabel={Comp. Time (s)},
		xmin=4.8, xmax=8.6,ymin=0.05, ymax=40,
		xtick={5.33,6.19,7.14,8.21},ytick={0.1,1.0,10},scaled ticks=false,tick label style={/pgf/number format/fixed},log ticks with fixed point,xmajorgrids=true, xminorgrids=true,ymajorgrids=true,yminorgrids=true,grid style=dashed]
		\addplot[smooth,color=NL1,thick,mark=*] table [x=GG,y=NL1_TIME]{Files/AGGRESSIVE.dat};	
		\addplot[smooth,color=NL5,thick,mark=*] table [x=GG,y=NL5_TIME]{Files/AGGRESSIVE.dat};	
		\addplot[smooth,color=RRMP,thick,mark=*] table [x=GG,y=RRMP_TIME]{Files/AGGRESSIVE.dat};	
		\addplot[smooth,color=RBMP,thick,mark=*] table [x=GG,y=RBMP_TIME]{Files/AGGRESSIVE.dat};
		\addplot[smooth,color=TRMP,thick,mark=*] table [x=GG,y=TRMP_TIME]{Files/AGGRESSIVE.dat};	
		\addplot[smooth,color=TBMP,thick,mark=*] table [x=GG,y=TBMP_TIME]{Files/AGGRESSIVE.dat};
		\addplot[semithick,dash pattern=on 2mm off 1mm,violet] coordinates{(8.3,0.01) (8.3,40)} node [pos=0.4,anchor=north, font=\footnotesize, sloped] {$\mu g$};		
	\end{semilogyaxis}
\end{tikzpicture}
\caption{Computational performance of the nonlinear and hybrid MPC controllers during lane change maneuver in the friction offset scenario for different levels of aggressive steering in the g--g diagram.}\label{fig:aggressive}
\end{center}
\end{figure}

For a more clear understanding of the computational performance during hazardous scenarios, we perform a second test in the friction offset case. Here, we fix the road friction to $\mu = 0.85$ and compare the computational performance of the controllers during the safe lane change maneuver over a range of steering actions to get closer to the boundary of the g--g diagram in Fig.~\ref{fig:manstab}. Figure~\ref{fig:aggressive} shows the tracking errors and computation time for MILP and NLP controllers for different levels of aggressiveness in terms of the acceleration magnitude $\sqrt{a_x^2+a_y^2}$, which is bounded by $\mu g$.

The computational efficiency of hybrid MPC can be easily seen in Fig.~\ref{fig:aggressive}. The closer the maneuver gets to the boundary of the stability constraint in the g--g diagram, the higher the tracking errors get for all the controllers. However, the error increase for the NLP controllers is much higher as they fail to converge to an acceptable optimum with 1 or even 5 starting points. Therefore, when the acceleration magnitude exceeds 6.2 m/s$^2$, even the NL--5 controller reaches higher mean and maximum tracking errors compared to the T--BMP, R--BMP, and R--RMP controllers. This is happening while NL--5 requires about 20 times more time to converge to its final solution. 

Comparing among the MILP controllers, T--BMP is the best hybrid MPC controller as it shows the most steady computation time and the lowest increase of tracking error as the steering action gets more extreme in our simulations.

\section{IPG CarMaker Simulation}\label{sec:ipg}

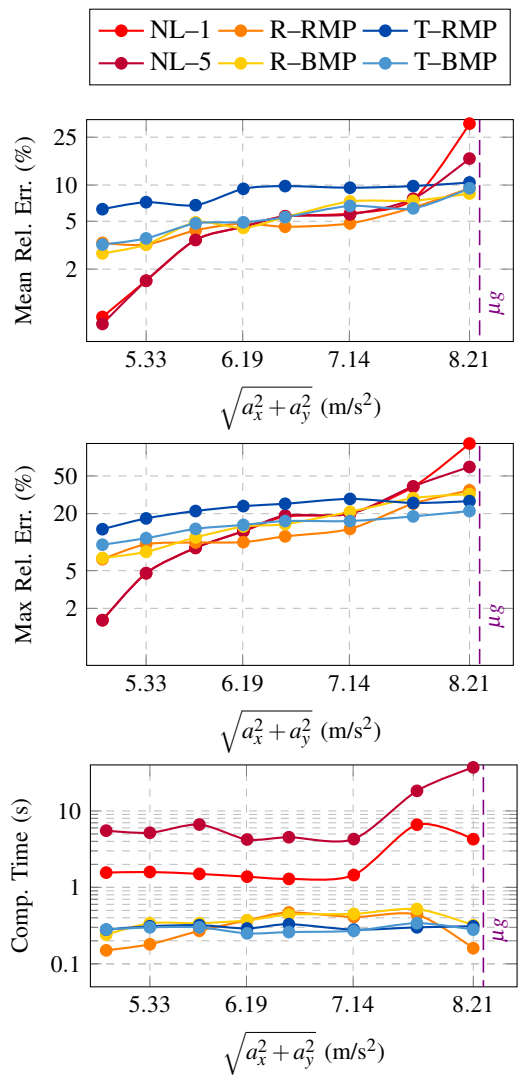
\begin{figure*}[hbt]
	\begin{center}
		\begin{tikzpicture}
			\begin{axis}[width=0.4\textwidth,height=0.25\textwidth,	
				xlabel={$t$ (s)},ylabel={$v_x$ (m/s)},
				xmin=0, xmax=5,ymin=31, ymax=38,
				xmajorgrids=true,xminorgrids=true,ymajorgrids=true,yminorgrids=true,grid style=dashed,legend pos=outer north east]	
				\addplot[smooth,color=black,very thick] table [x=T,y=VXREF]{Files/IPG_RRMP.dat};	
				\addplot[smooth,color=RRMP,very thick] table [x=T,y=IPGVX]{Files/IPG_RRMP.dat};	
				\addplot[smooth,color=RBMP,very thick] table [x=T,y=IPGVX]{Files/IPG_RBMP.dat};	
				\addplot[smooth,color=TRMP,very thick] table [x=T,y=IPGVX]{Files/IPG_TRMP.dat};
				\addplot[smooth,color=TBMP,very thick] table [x=T,y=IPGVX]{Files/IPG_TBMP.dat};		
				\legend{Reference,R--RMP, R--BMP, T--RMP, T--BMP}
			\end{axis}
		\end{tikzpicture}
		\begin{tikzpicture}
			\begin{axis}[width=0.4\textwidth,height=0.25\textwidth,	
				xlabel={$t$ (s)},ylabel={$\delta$ (rad)},
				xmin=0, xmax=5,%ymin=31, ymax=38,
				xmajorgrids=true,xminorgrids=true,ymajorgrids=true,yminorgrids=true,grid style=dashed,legend pos=outer north east]	
				\addplot[smooth,color=RRMP,very thick] table [x=T,y=DELTAREF]{Files/IPG_RRMP.dat};	
				\addplot[smooth,color=RBMP,very thick] table [x=T,y=DELTAREF]{Files/IPG_RBMP.dat};	
				\addplot[smooth,color=TRMP,very thick] table [x=T,y=DELTAREF]{Files/IPG_TRMP.dat};
				\addplot[smooth,color=TBMP,very thick] table [x=T,y=DELTAREF]{Files/IPG_TBMP.dat};		
				\legend{R--RMP, R--BMP, T--RMP, T--BMP}
			\end{axis}
		\end{tikzpicture}
		\begin{tikzpicture}
			\begin{axis}[width=0.25\textwidth,height=0.25\textwidth,	
				xlabel={$a_x$ (m/s$^2$)},ylabel={$a_y$ (m/s$^2$)},
				xmin=-10, xmax=10,ymin=-10, ymax=10,
				xmajorgrids=true,xminorgrids=true,ymajorgrids=true,yminorgrids=true,grid style=dashed,legend pos=outer north east]	
				\draw[color=red,fill=none,thick](0,0) circle (8.5);
				\filldraw[color=gray,fill=red,fill opacity=0.1,thick,dashed](0,0) circle (8.0);
				\filldraw[color=gray,fill=green,fill opacity=0.1,thick,dashed](0,0) circle (4.25);
				\addplot[smooth,color=RRMP,very thick] table [x=IPGAX,y=IPGAY]{Files/IPG_RRMP.dat};	
				\addplot[smooth,color=RBMP,very thick] table [x=IPGAX,y=IPGAY]{Files/IPG_RBMP.dat};	
				\addplot[smooth,color=TRMP,very thick] table [x=IPGAX,y=IPGAY]{Files/IPG_TRMP.dat};
				\addplot[smooth,color=TBMP,very thick] table [x=IPGAX,y=IPGAY]{Files/IPG_TBMP.dat};		
				\legend{R--RMP, R--BMP, T--RMP, T--BMP}
			\end{axis}
		\end{tikzpicture}
		\begin{tikzpicture}
			\begin{axis}[width=0.25\textwidth,height=0.25\textwidth,	
				xlabel={$\beta$ (rad)},ylabel={$r$ (rad/s)},tick label style={/pgf/number format/fixed},xtick={-0.1,0,0.1},
				xmin=-0.15, xmax=0.15,ymin=-0.6, ymax=0.6,
				xmajorgrids=true,xminorgrids=true,ymajorgrids=true,yminorgrids=true,grid style=dashed,legend pos=outer north east]	
				\addplot[smooth,color=RRMP,very thick] table [x=IPGBETA,y=IPGR]{Files/IPG_RRMP.dat};	
				\addplot[smooth,color=RBMP,very thick] table [x=IPGBETA,y=IPGR]{Files/IPG_RBMP.dat};	
				\addplot[smooth,color=TRMP,very thick] table [x=IPGBETA,y=IPGR]{Files/IPG_TRMP.dat};
				\addplot[smooth,color=TBMP,very thick] table [x=IPGBETA,y=IPGR]{Files/IPG_TBMP.dat};		
			\end{axis}
		\end{tikzpicture}
		\begin{tikzpicture}
			\begin{axis}[width=0.25\textwidth,height=0.25\textwidth,	
				xlabel={$\delta$ (rad)},ylabel={$v_y$ (m/s)},xtick={-0.15,0,0.15},
				xmin=-0.2, xmax=0.2,ymin=-4.5, ymax=2.5,
				xmajorgrids=true,xminorgrids=true,ymajorgrids=true,yminorgrids=true,grid style=dashed,legend pos=outer north east]	
				\addplot[smooth,color=RRMP,very thick] table [x=DELTAREF,y=IPGVY]{Files/IPG_RRMP.dat};	
				\addplot[smooth,color=RBMP,very thick] table [x=DELTAREF,y=IPGVY]{Files/IPG_RBMP.dat};	
				\addplot[smooth,color=TRMP,very thick] table [x=DELTAREF,y=IPGVY]{Files/IPG_TRMP.dat};
				\addplot[smooth,color=TBMP,very thick] table [x=DELTAREF,y=IPGVY]{Files/IPG_TBMP.dat};		
				\legend{R--RMP, R--BMP, T--RMP, T--BMP}
			\end{axis}
		\end{tikzpicture}
		\caption{Computational performance of the nonlinear and hybrid MPC controllers during lane change maneuver in the friction offset scenario for different levels of aggressive steering in the g--g diagram.}\label{fig:ipg}
	\end{center}
\end{figure*}

To validate the performance of the MILP controllers, we have simulated the most aggressive double-lane change maneuver in handling-limit scenario using a high-fidelity BMW model in IPG CarMaker with the RealTime (RT) Tire model. Figure~\ref{fig:ipg} shows the state and input plots, as well as the tire force and constraint bounds from the simulations. Despite the fact that the MILP controllers use a much simpler prediction model and friction offset, comparing the states shows that they are able to provide control inputs that satisfy the constraint boundaries in the handling limits as shown on the g--g diagram.

\section{Discussion and Outlook}\label{sec:discuss}% ~ ~ ~ ~ ~ ~ ~ ~ ~ ~ ~ ~ ~ ~ ~ ~ ~ ~ ~ ~ ~ ~ ~ ~ ~ ~ ~ ~ ~ ~ ~ ~ 

\subsection{MILP vs.\ MIQCP}
In general, as MILPs are solved faster than MIQCPs, MILP-based approaches are more suitable choices in terms of computation speed. The control performance highly depends on the accuracy of the hybrid approximation. Consequently, the tracking performance of MIQCP controllers was not as good as MILP controllers due to their less accurate constraint approximation compared to their MILP counterparts. Nevertheless, for systems or applications where a lower approximation error with mixed-integer quadratic constraints is obtained for the nonlinear constraints, MIQCP can be an efficient formulation to hybridize the nonlinear MPC problem. This can for instance be the case for systems with quadratic or bi-variate nonlinearities where considering the quadratic terms helps avoid using large number of local linear or affine modes to obtain the same level of accuracy.

\subsection{Robustness to Friction Uncertainty}

When friction uncertainty is present, MILP controllers do not require longer computation times, whereas NLP controllers take significantly longer to converge. In addition, the rate of increase in tracking error for the NLP controllers is higher than that of the MILP controllers in the presence of a friction offset.

Furthermore, MILP controllers are better able to recover from friction disturbances due to their more comprehensive search for an optimal solution in the decision space and the fact that they always reach the global optimum if given enough time. This means that if the error is already very high in the initial state for the current time step, the NLP controllers may not find a feasible solution at all in the vicinity of their initial guesses, while the MILP controller will converge to one via branch-and-bound strategy, provided that they have given sufficient time. As a result, even in cases where both nonlinear and hybrid controllers recover from high tracking errors during the friction disturbance, the MILP controller reaches smaller tracking error in fewer control time steps.

The robustness of the MILP controllers compared to the NLP controllers is summarized in Fig.~\ref{fig:comp}. While NLP controllers can reach lower relative errors, their behavior in terms of accuracy, as well as computation speed, is not as consistent as MILP controllers in the presence of friction offset or disturbance. However, MILP controllers are considerably (at least 10 times) quicker to converge to their optimal solution and show significantly less variations in the relative error when exposed to friction uncertainty.

\subsection{Performance Close to Handling Limits}

Getting closer to the handling limits leads to larger tracking errors for all the controllers. However, NLP controllers may deviate from the reference significantly as they may fail to converge to an acceptable optimum. Meanwhile, the MILP controllers converge to lower errors in a shorter time, e.g.,\ in the boundary of the acceleration magnitude, the best MILP controller converges to 30\% of the tracking error of the best NLP controller with 95\% reduction of its corresponding computation time. This shows that MILP controllers are more suitable choices for real-time control in emergency evasive maneuvers.

\subsection{Overall Computational Performance}

The shortcoming of a warm-start strategy in solving an NLP is more clear in emergency maneuvers, which stems from its limitation in searching for the optimal solution in the search space during aggressive maneuvers. In addition, even with a multi-start strategy, the NLP controller may converge to worse solutions if the uncertainty influences the feasible region and the NLP solutions become infeasible for the real system. However, the branch-and-bound approach of the MILP solver has a better exploration of possible solutions than a potentially real-time implementable NLP solver such as gradient-based solver, while keeping the computation time as low as for the ideal case.

Even in the ideal case where the NLP controllers benefit from employing the same prediction model as the real system, they show high variations in tracking control and computation time, which means they are not suitable options for robust control in hazardous scenarios. In the presence of uncertainty, the NLP controllers suffer from larger tracking errors as well as an exponential growth in their computation time. However, the MILP controllers converge to smaller tracking errors, within a much smaller variation bound, and with significantly less computation times.

\section{Conclusions}\label{sec:conc}% ~ ~ ~ ~ ~ ~ ~ ~ ~ ~ ~ ~ ~ ~ ~ ~ ~ ~ ~ ~ ~ ~ ~ ~ ~ ~ ~ ~ ~ ~ ~ ~ 

This paper has presented a comparative assessment of nonlinear MPC controllers vs.\ their various hybridized counterparts in terms of computational efficiency for vehicle control during emergency evasive maneuvers. The hybridization of the nonlinear problem was presented and discussed in Part I of this publication, where several guidelines for hybridization are given in a generalized framework. 

The benchmark of this paper uses three hybridized models and four hybridized constraint formulations for a nonlinear single track vehicle model considering nonlinear physics-based constraints for stability and tire-force saturation. Five reference maneuvers were selected to represent emergency situations where the computational efficiency is crucial for real-time proactive vehicle control. The hybrid and nonlinear controllers then were compared in multiple scenarios to compare their control performance and computation time, and their robustness in the presence of friction uncertainty in the form of an offset or a disturbance. Further, we studied the tracking behavior of the controllers with respect to how close the vehicle is operating in handling limits. The conclusions of our comparative assessment are summarized next with respect to different criteria.

Based on our comparative assessment, combining hybrid MPC and hybrid predictive estimation techniques (e.g.,\ moving-horizon estimation) is an important next research step for improving robustness in hazardous driving scenarios. Moreover, as quadratic forms of nonlinearity are extensively encountered in modeling of physical systems, we will also investigate piecewise-quadratic-based hybridization of the prediction model and physics-based constraints for MIQCP formulation of MPC optimization problem. This can particularly be beneficial for systems with nonlinearities that can better be approximated using quadratic approximations and can lead to significant improvements in terms of accuracy and computational efficiency of the hybrid controller. In addition, we will extend the prediction model to include the effects of wheel-speed/torque dynamics for improved control performance in hazardous scenarios. Finally, we suggest hardware implementation of the MILP tracking controllers is also a very relevant topic for future research.

%\vspace{0.75cm}

%All the above mentioned factors make the MILP controllers a good choice for real-time implementation of MPC in emergency situations. 

\section*{Acknowledgment}\label{sec:ack}% ~ ~ ~ ~ ~ ~ ~ ~ ~ ~ ~ ~ ~ ~ ~ ~ ~ ~ ~ ~ ~ ~ ~ ~ ~ ~ ~ ~ ~ ~ ~ ~ 

This research is funded by the Dutch Science Foundation NWO-TTW within the EVOLVE project (no.\ 18484) and by the European Research Council (ERC) under the European Union’s Horizon 2020 research and innovation programme within the CLariNet project (no.\ 101018826).

\appendices

\section{Nonlinear System Description}\label{app:model}

A single-track nonlinear representation of the vehicle model is described by the following equations~\cite{Chowdhri2021}: 
\begin{equation}
	\dot{v}_x = \frac{1}{m} \left[F_{x\rm{f}} \cos{\delta} - F_{y\rm{f}} \sin{\delta} + F_{x\rm{r}}\right]+v_{y}r,
	\label{eq:vxnonlin}
\end{equation}
\begin{equation}
	\dot{v}_y = \frac{1}{m} \left[F_{x\rm{f}} \sin{\delta} + F_{y\rm{f}} \cos{\delta} + F_{y\rm{r}}\right]-v_{x}r,
	\label{eq:vynonlin}
\end{equation}
\begin{equation}
	\dot{r} = \frac{1}{I_{zz}} \left[F_{x\rm{f}} \sin{\delta} \; l_\text{f} + F_{y\rm{f}} \cos{\delta} \; l_\text{f} - F_{y\rm{r}} \;l_\text{r} \right],
	\label{eq:rnonlin}
\end{equation}
and the lateral forces are given by the Dugoff model~\cite{Dugoff1970}
\[ F_{ya} = \dfrac{C_{\alpha_{a}}}{1-\kappa_{a}} f_\lambda(\lambda^w_{a}) \alpha_{a},\]
with $a \in \{\rm{f}, \rm{r}\}$ where $\mu_a$ is the varying friction coefficient, and $\lambda^w_{a}$ and $f_\lambda$ are the weighting coefficient and function, defined as
\[\mu_a = \mu_0 \left(1-e_{\rm{r}} v_x \sqrt{\kappa_{a}^2 +\tan^2{{\alpha_{a}}}}\right),\]
\[\lambda^w_{a} = \frac{\mu_a F_{z{a}} (1-\kappa_{a})}{2 \sqrt{(C_{\kappa_{a}} \kappa_{a})^2+(C_{\alpha_{a}} \tan{\alpha_{a}})^2}},\]
\[f_\lambda(\lambda^w_a) =    
\begin{cases} 
	\lambda^w_{a}(2-\lambda^w_{a}) & \lambda^w_{a} < 1 \\
	1 & \lambda^w_{a} \geq 1
\end{cases}.\]
We selected the Dugoff tire with varying friction to capture nonlinear behavior of the tire with respect to changes in friction, normal force, slip angle, velocity, etc. This model captures the decay in force after saturation~\cite{Chowdhri2021} and is used in the literature~\cite{Mirzaeinejad2010}, also for extreme maneuvers~\cite{Baars2021,Zhang2022}. The feasible region is defined by two other physics-based constraints: the working limits of the vehicle (known as the g-g diagram constraint~\cite{Chowdhri2021})
\begin{equation}
	\left(\dot{v}_x - v_y r\right)^2 + \left(\dot{v}_y + v_x r\right)^2 \leqslant (\min \{\mu_{\rm{f}}, \mu_{\rm{r}}\}\; g)^2,
	\label{eq:gg}
\end{equation}
and the saturation limits of the tires known as the Kamm circle constraint~\cite{Chowdhri2021},
\begin{equation}
	F_{xa}^2 + F_{ya}^2 \leqslant (\mu_a F_{za})^2, \quad a \in \{\rm{f}, \rm{r}\}.
	\label{eq:kamm}
\end{equation}
\begin{table}[htbp]
	\caption{System variables}
	\begin{center}
		\begin{tabular}{c|c|c|c}
			\hline
			\textbf{Var.} & \textbf{Definition} & \textbf{Unit} & \textbf{Bounds}\\
			\hline
			$v_x$ & Longitudinal velocity & m/s & [5, 50]\\
			$v_y$ & Lateral velocity & m/s & [-10, 10]\\
			$\psi$ & Yaw angle & rad & -- \\
			$r$ & Yaw rate & rad/s & [-0.6, 0.6]\\
			$\delta$ & Steering angle (road)& rad & [-0.5, 0.5]\\
			$F_{x\rm{f}}$ & Longitudinal force on the front axis & N & [-5000, 0]\\
			$F_{x\rm{r}}$ & Longitudinal force on the rear axis & N & [-5000, 5000]\\
			$F_{y\rm{f}}$ & Lateral force on the front axis & N & --\\
			$F_{y\rm{r}}$ & Lateral force on the rear axis & N & --\\
			$F_{z\rm{f}}$ & Normal load on the front axis & N & --\\
			$F_{z\rm{r}}$ & Normal load on the rear axis & N & --\\
			$\alpha_{\rm{f}}$ & Front slip angle & rad & --\\
			$\alpha_{\rm{r}}$ & Rear slip angle & rad & --\\
			$\kappa_{\rm{f}}$ & Front slip ratio & -- & --\\
			$\kappa_{\rm{r}}$ & Rear slip ratio & -- & --\\
			$\mu_{\rm{f}}$ & Friction coefficient on the front tire & -- & --\\
			$\mu_{\rm{r}}$ & Friction coefficient on the rear tire & -- & --\\
			\hline
			$x$ & State vector $\coloneqq \begin{bmatrix}
				v_x & v_y & r
			\end{bmatrix}^T$ & -- & -- \\
			$u$ & Input vector $\coloneqq \begin{bmatrix}
				F_{x\rm{f}} & F_{x\rm{r}} & \delta
			\end{bmatrix}^T$ & -- & -- \\
			\hline
		\end{tabular}
		\label{tab:vars}
	\end{center}
\end{table}

\begin{table}[htbp]
	\caption{System parameters$^\ast$}
	\begin{center}
		\begin{tabular}{c|c|c|c}
			\hline
			\textbf{Par.} & \textbf{Definition} & \textbf{Value} & \textbf{Unit}\\
			\hline
			$m$ & Vehicle mass & 1970 & kg\\
			$I_{zz}$ & Inertia moment about z-axis & 3498& kg/m$^2$\\
			$l_\text{f}$ & CoG$^{\ast\ast}$ to front axis distance& 1.4778 & m\\
			$l_\text{r}$ & CoG to rear axis distance & 1.4102 & m\\
			$C_{\alpha_\text{f}}$ & Front cornering stiffness &126784 & N \\
			$C_{\alpha_\text{r}}$ & Rear cornering stiffness & 213983 & N\\
			$C_{\kappa_\text{f}}$ & Front longitudinal stiffness & 315000 & N \\
			$C_{\kappa_\text{r}}$ & Rear longitudinal stiffness & 286700 & N\\
			$\mu_0$ & Zero-velocity friction & 1.076 & --\\
			$e_{\rm{r}}$ & Friction slope & 0.01 & --\\
			\hline
			\multicolumn{4}{l}{$^{\ast}$These values correspond to the IPG CarMaker BMW vehicle model}\\
			\multicolumn{4}{l}{$^{\ast\ast}$Center of Gravity}
		\end{tabular}
		\label{tab:params}
	\end{center}
\end{table}

\bibliographystyle{IEEEtran}
\bibliography{Citations}

\vspace*{-1cm}
\begin{IEEEbiography}[{\includegraphics[width=1in,height=1.25in,clip,keepaspectratio]{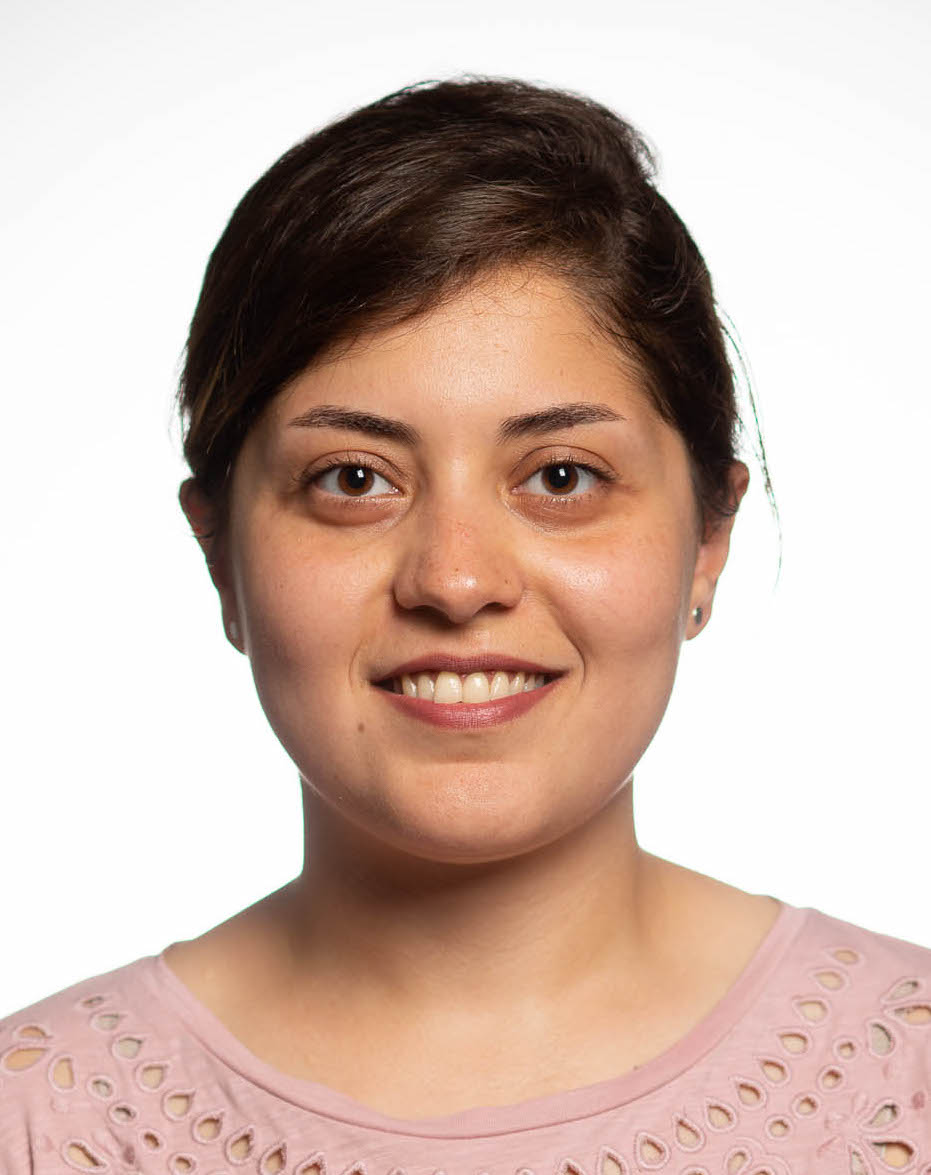}}]{Leila Gharavi} 
	is a PhD candidate at Delft Center for Systems and Control, Delft University of Technology, The Netherlands. She received her BSc and MSc degrees in mechanical engineering from Amirkabir University of Technology (Tehran Polytechnic) in Iran and has research experience in automatic manufacturing and production, vibration analysis and control of nonlinear dynamics, and soft rehabilitation robotics. 
	
	Currently, her research focuses on nonlinear and hybrid systems, optimization, and model-predictive control, with applications to adaptive and proactive control of automated  vehicles in hazardous scenarios.
\end{IEEEbiography}
\vspace*{-2cm}
\begin{IEEEbiography}[{\includegraphics[width=1in,height=1.25in,clip,keepaspectratio]{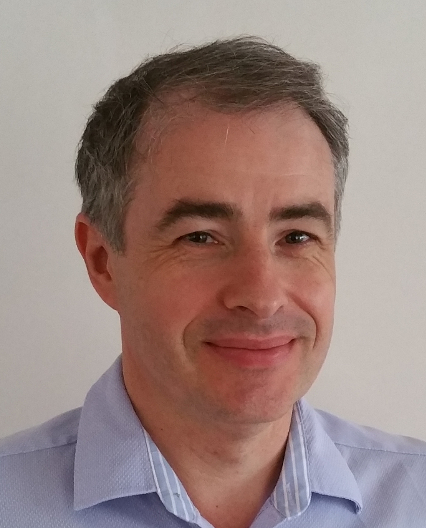}}]{Bart De Schutter}(Fellow, IEEE) 
	received the PhD degree (\emph{summa cum laude}) in applied sciences from KU Leuven, Belgium, in 1996. He is currently a Full Professor and Head of Department at the Delft Center for Systems and Control, Delft
	University of Technology, The Netherlands. His research interests include multi-level
	and multi-agent control, model predictive control, learning-based control, and control
	of hybrid systems, with applications in intelligent transportation systems and smart energy systems. 
	
	Prof.\ De Schutter is a Senior Editor of the IEEE Transactions on Intelligent Transportation Systems.
\end{IEEEbiography}
\vspace*{-2cm}
\begin{IEEEbiography}[{\includegraphics[width=1in,height=1.25in,clip,keepaspectratio]{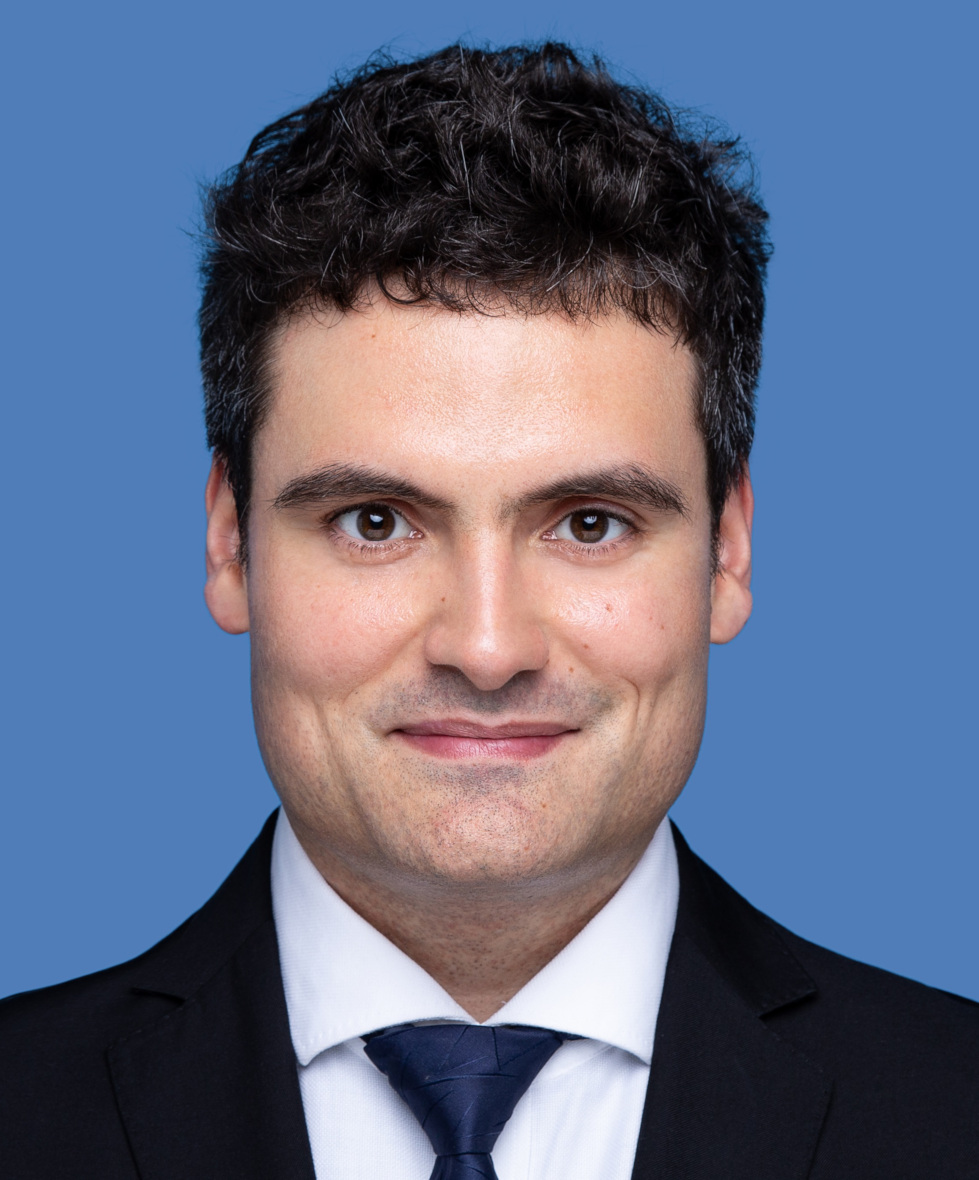}}]{Simone Baldi}
	(Senior Member, IEEE) received the B.Sc. in electrical engineering, and the M.Sc. and Ph.D. in automatic control engineering from University of Florence, Italy, in 2005, 2007, and 2011, respectively. Since 2019, he is a Professor with Southeast University, China, with a guest position with Delft Center for Systems and Control, Delft University of Technology, The Netherlands, where he was Assistant Professor in 2014-2019. His research interests include adaptive and learning systems with applications in intelligent vehicles and smart energy. He was awarded outstanding Reviewer of Applied Energy in 2016, Automatica in 2017, AIAA Journal of Guidance, Control, and Dynamics in 2021. He is a Subject Editor of International Journal of Adaptive Control and Signal Processing, a Technical Editor of IEEE/ASME Transactions on Mechatronics, and an Associate Editor for IEEE Control Systems Letters and Journal of the Franklin Institute.
\end{IEEEbiography}

\end{document}